\documentclass[fleqn,usenatbib]{mnras}

\usepackage{newtxtext,newtxmath}
\usepackage[T1]{fontenc}

\DeclareRobustCommand{\VAN}[3]{#2}
\let\VANthebibliography\thebibliography
\def\thebibliography{\DeclareRobustCommand{\VAN}[3]{##3}\VANthebibliography}

\usepackage{graphicx}	% Including figure files
\usepackage{amsmath}	% Advanced maths commands

\usepackage[dvipsnames]{xcolor}

\newcommand{\vecF}{\mbox{\boldmath $F$} {}}
\newcommand{\vecr}{\mbox{\boldmath $r$} {}}

\newcommand{\vece}{\mbox{\boldmath $e$} {}}
\newcommand{\vecv}{\mbox{\boldmath $v$} {}}

\title[Eccentric bodies in a ringed disc]{Orbital evolution of highly-eccentric bodies embedded in a ringed accretion disc} 
\author[R. A. Anaya-S\'anchez and F. J. S\'anchez-Salcedo]{R. A. Anaya-S\'anchez
and F. J. S\'anchez-Salcedo\thanks{E-mail: jsanchez@astro.unam.mx} \\
Universidad Nacional Aut\'onoma de M\'exico, Instituto de Astronom\'{\i}a, 
A. P. 70-264,  04510 Ciudad de M\'exico, Mexico\\
}

\date{Accepted XXX. Received YYY; in original form ZZZ}

\pubyear{2024}

\begin{document}
\label{firstpage}
\pagerange{\pageref{firstpage}--\pageref{lastpage}}
\maketitle

\begin{abstract}
Various processes can induce long-lived overdense rings and arcs in
protoplanetary and AGN accretion discs, such as the accumulation of gas at the outer
edge of the dead zone, or the infall of material. Using the local approximation of
dynamical friction, we investigate the orbital evolution of a
low-mass highly-eccentric point-mass accretor (perturber) embedded in an 
isothermal disc hosting a density ring.
We specifically consider the regime in which the eccentricity exceeds four times the disc aspect ratio.
For prograde perturbers, orbits that cross the ring progressively
circularize while their semi-major axes converge toward the ring radius.
As a result, perturbers accumulate, forming a population ring superimposed 
on the gaseous ring. The ring therefore acts as a migration trap  
for these eccentric orbits. We also find that prograde orbits 
tangent to the ring, either at apocentre or pericentre, remain tangential throughout their 
evolution; perturbers confined to these trajectories experience the 
highest accretion rates. 
%For retrograde perturbers, all orbits drift toward the central object. 
In contrast, retrograde perturbers always migrate inward. Once the semi-major axis becomes smaller 
than the ring radius, the eccentricity grows, but not enough for the orbit to intersect the ring again.
We also discuss how feedback effects, such as jet launching and
thermal torques, could modify the effective forces acting on the perturbers.

\end{abstract}

\begin{keywords}
accretion, accretion discs -- galaxies: active -- planetary systems: formation -- planetary systems: protoplanetary discs -- planet-disc interactions.
\end{keywords}

% Keywords must be from the standard list and in alphabetical order. 

%%%
%%% Beginning of document proper
%%%

\section{Introduction}

The orbital evolution of a gravitational body embedded in a gaseous disc has significant
applications in astrophysics, ranging from the dynamics of protoplanets or planetary cores
in the protoplanetary discs to the inspiraling of compact objects (stars, stellar remnants
and black holes) in the accretion discs of AGNs. The orbiter-disc interaction enables
the exchange of angular momentum, driving the radial migration of the orbiter
and changing its orbital eccentricity.

In protoplanetary discs, long-lived gas rings are expected to emerge at the transition region 
between magnetically inactive dead zones and a region where the MRI operates \citep[e.g.,][]{var06,lyr09}. In fact, 
MHD simulations by \citet{pin16} shows that such rings can reach 
density contrasts of up to a factor of $5$, and persist for at least $1$ Myr \citep[see also][]{kad19}. In addition, variations in the depth of the active
layer across the snowline could also promote the formation of a density bump \citep{kre07}.
The resulting pressure bumps can halt the radial drift of dust particles, naturally leading to
the dust rings and gaps revealed in high-resolution interferometric observations \citep[e.g.,][]{and18,dul18,lon18,cie21}.

Protoplanetary discs and star-forming regions are not isolated systems
but interact with the surrounding natal cluster and, therefore, they may be subject to
external perturbations \citep[e.g.,][]{moe09}. \citet{vor15,vor16} show that the infall 
of material has a crucial impact on the evolution of circumstellar discs, to the extent that discs 
formed in the cores of clouds can have an external counter-rotating disc (separated by
a gap), if the external environment was counter-rotating with respect to the core.

On the other hand, protostars may undergo late accretion via capture or infall events of 
cloudlets of gas, forming second generation discs \citep[e.g.,][]{dul19}.
\citet{kuf20} show that the accretion of a cloudlet could lead to density
bumps, spiral arms or arc-like structures as those observed around AB Aurigae and
around HD 1000546. 

Radio interferometry observations have revealed the existence
of streamers that are probably falling onto planet-forming discs and star-forming
systems \citep[e.g.][]{cac24}. The accretion streamers could trigger
the formation of spiral waves \citep{thi11,hen17}, shocks
\citep{gar22} and induce the formation of rings and
gaps \citep{bae15,kuz22}. Interestingly, \citet{zha25} suggest that
pressure bumps arising from gas infall could stop the inward drifting motion of the dust, promoting planetesimal formation through the
streaming instability.

In the context of galactic nuclei, giant molecular clouds may be tidally disrupted by the central massive black hole, forming eccentric, multi-ringed discs \citep{bon08}. If sufficiently massive, these rings may fragment and form stars.

AGN discs can be also rejuvenated through gas accretion. Gas replenishment is
expected to occur after a major merger with a rich-gas galaxy. It is likely that the 
merger will result in black hole binary and a turbulent environment in which cold 
clumps may be continuously falling to the galactic centre forming a circumbinary disc
with ringed and arc-like structures \citep{dun14,goi16,goi17,mau18}.

In regions of the disc with a positive radial density gradient, as well as in cavities or pressure bumps, the torques acting on embedded bodies, and hence their migration, can
be substantially modified \citep{mas06,ogi10,mut11,duq25}. 
For circular orbits and in the region where the disc has a positive density gradient, 
a trapping radius can arise where the corotation torque balances the Lindblad torque
\citep{mas06, rom19, cha24}. In such cases, perturbers migrating from either larger 
or smaller radii than the trapping radius tend to converge toward it.
\citet{mut11} investigated how the timescales of radial migration and eccentricity damping  of eccentric planets depend on surface density radial profile, assuming $\Sigma(R) \propto R^{-\beta}$, including cases with $\beta<0$.

The observation of dust rings in protoplanetary discs has revived the long-standing
problem of planet and embryo migration in structured discs. 
In such ringed discs, migration cannot be understood solely in terms of classical 
Lindblad and corotation torques. In particular, thermal torques arising from the heat released by 
ongoing dust accretion onto the embryos or planets may alter the density structure around them, thereby modifying migration rates and influencing trapping \citep[e.g.,][]{chr23,pie24,vel24}. These studies have concentrated on embryos with eccentricities
of the order of, or smaller than, the disc’s aspect ratio $h$.

In the present work, we investigate the orbital evolution of highly eccentric bodies ($e\geq 4h$, where $h$ is the disc aspect ratio), embedded in ringed isothermal discs, under both prograde and retrograde configurations. We focus on bodies of sufficiently low mass that they do not carve a gap in the disc.
For such high eccentricities, the perturbers move supersonically, which allows us to apply dynamical friction and accretion drag approximations. 
We first compute the orbital evolution in a purely gaseous accretion disc, treating the perturber as a point-mass accretor.
We then examine 
the impact of additional effects (such as thermal torques or jet launching) that can modify the
structure of the wake around the perturber. Our results are relevant for both stellar-mass compact objects in discs around
supermassive black holes and for eccentric embryos and planetary cores in ringed protoplanetary discs.

The  paper is organized as follows. In Section \ref{sec:model} we introduce the model
and underlying assumptions, and establish the framework for describing the orbital evolution 
of a perturber embedded in the disc. Section \ref{sec:results} presents and interprets the results of 
our analysis. In Section \ref{sec:applicability}, we discuss the applicability of our
results to astrophysical systems by examining additional physical processes, such as thermal feedback.
Finally, Section \ref{sec:summary} summarizes the main outcomes of this study.

\begin{table}
  \caption{Parameters of the initial profiles of the discs.}
  \label{tab:parameters}
  \begin{tabular}{llllr}\hline
   
    & \multicolumn{1}{c}{$\beta$} 
    & \multicolumn{1}{c}{$\Sigma_{\rm bg}$} 
    & \multicolumn{1}{c}{$M_{\rm ring}$} 
    & \multicolumn{1}{c}{$\tau_{i}$} \\
    & 
    & \multicolumn{1}{c}{[$M_{\bullet}/R_{0}^{2}$]} 
    & \multicolumn{1}{c}{[$M_{\bullet}$]} 
    & \\
    \hline
    Models 1A--1C & $1$   & $2\times 10^{-3}$ & $0.03$  & $0.018$ \\
    Models 2A--2C & $0.5$ & $10^{-4}$         & $0.005$ & $0.004$ \\
    \hline
  \end{tabular}
\end{table}

\begin{table}
  
  \caption{Ring spreading timescale, $t_{\nu}$, in units of $(2\pi/\omega_{0})$.} 
 \label{tab:tnu}
 \begin{tabular}{llll}\hline
   
     & \multicolumn{1}{c}{A} & \multicolumn{1}{c}{B} & \multicolumn{1}{c}{C} \\
  
    Model 1   & $1.5\times 10^{4}$ & $3\times 10^{3}$ & $15.2$ \\
    Model 2 & $3.2\times 10^{3}$ & $640$ & $6.4$ \\
 
    \hline
    
  \end{tabular}
\end{table}

\section{Model and assumptions}
\label{sec:model}
\subsection{The disc}
We consider an axisymmetric gaseous disc composed of a steady background component with a power-law surface density profile, and a viscously spreading ring. Specifically, the gas surface density of the disc is parameterized as follows:
\begin{equation}
\Sigma(x,\tau)=\frac{\Sigma_{\rm bg}}{x^{\beta}}+\left(\frac{M_{\rm ring}}{\pi R_{0}^{2}}\right)\frac{1}{(\tau +\tau_{i}) x^{1/4}}
I_{1/4}\left(\frac{2x}{\tau+\tau_{i}}\right) \exp\left(-\frac{1+x^{2}}{\tau+\tau_{i}}\right),
\label{eq:Sigma}
\end{equation}
where $x\equiv R/R_{0}$, with $R_{0}$ is the radius where the ring is centred initially, $\tau\equiv 12\nu t/R_{0}^{2}$ (with $\nu$ the viscosity), $\Sigma_{\rm bg}$ is the surface density of the background disc at $R=R_{0}$, 
$M_{\rm ring}$ is the mass of the ring, and $I_{1/4}$
is the modified Bessel function of the first kind and order $1/4$
\citep[e.g.,][]{lyn74, pri81, spe03, jos23}.  
Observe that setting $\tau_{i}=0$ results in a ring with zero thickness at $t = 0$. To start with a ring of finite thickness, we choose a nonzero value for $\tau_{i}$. Throughout this paper,  
$\omega_{0}$ denotes the angular frequency of the disc at $R=R_{0}$, that is, $\omega_{0}=\sqrt{GM_{\bullet}/R_{0}^{3}}$, with $M_{\bullet}$ the mass of the central object.

Note that the overdense ring is assumed to be axisymmetric, which implies that the ring is stable against the Rossby-wave instability (RWI). Since our primary focus is on the long-term evolution of eccentric bodies, we regard this approximation as acceptable.

\begin{figure*}
  \vspace{0pt}
  \includegraphics[angle=0,width=1\textwidth]{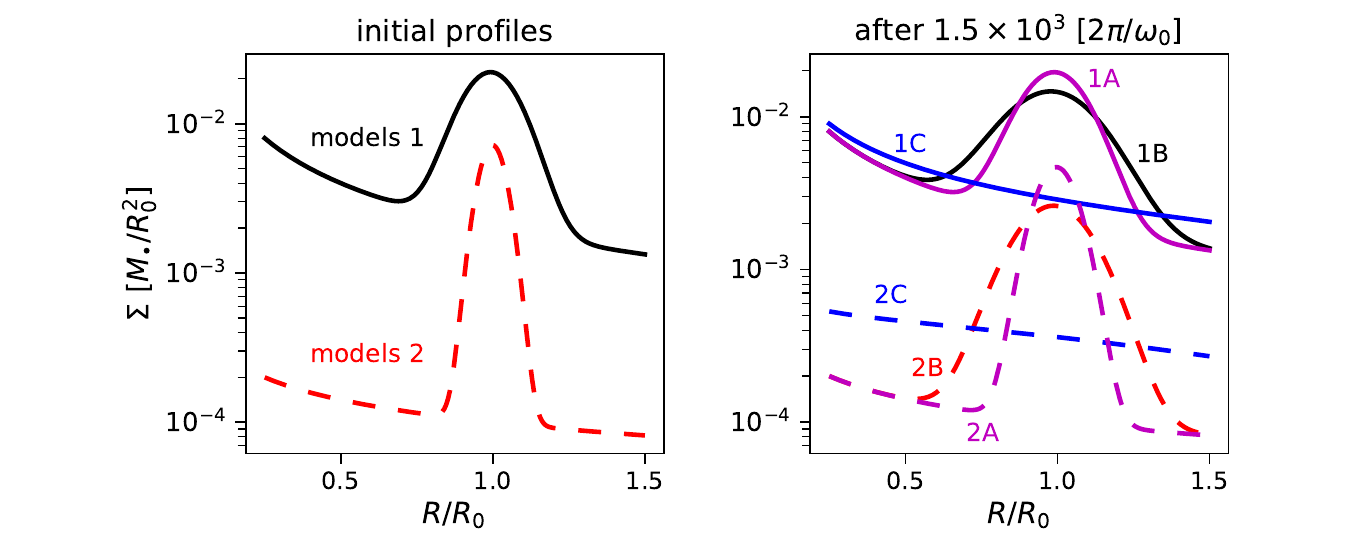}
  \caption{Radial profiles of the unperturbed surface density of the disc at 
  $t=0$ (left panel) and at $t=1.5\times 10^{3}(2\pi/\omega_{0})$ (right panel) for the different models (see Tables \ref{tab:parameters} and \ref{tab:tnu}). At $t=0$, models 1A-1C exhibit identical profiles, and likewise for models 2A-2C.}
  \label{fig:surface_density}
%This figure is created by mac: Documents/disco6/tex/Rodrigo_Anaya/viscous_disk_v2/surface_density/evolving_surface_density_v4.py
\end{figure*}

We explore six models, whose parameters are summarized in Table \ref{tab:parameters}.
The left panel of Figure \ref{fig:surface_density} shows their initial surface density profiles. Models 1A-1C have identical initial density profiles. The same applies to models 2A-2C. The distinction between models A, B and C lies in the adopted effective viscosity, which leads to different evolutionary outcomes (right panel of Figure \ref{fig:surface_density}). Models A use the lowest viscosity ($\nu=0.5\times 10^{-7}\omega_{0}^{2}R_{0}$), resulting in slower viscous diffusion. It represents a scenario where the density bump is sustained by internal processes, as 
expected in the outer edge of the dead zone \citep{var06,reg13,pin16}. On the other hand, models
B adopt a higher effective viscosity ($\nu=2.5\times 10^{-7}\omega_{0}^{2}R_{0}$), mimicking the evolution of a ring formed through infall \citep{zha25}. 
Models C assume the highest viscosity, $\nu=2.5\times 10^{-5}\omega_{0}^{2}R_{0}$, representative of values expected in AGN discs.
Finally, models 2A-2C differ from models 1A-1C in that they have a less massive background component and are intended to represent the accretion of a streamer that has been disrupted by tidal forces. 

It is useful to define the ring spreading timescale, $t_{\nu}$, as
the time required for the surface density at the ring's maximum to decrease to
the midpoint between its initial value and that of the background disc.  Table \ref{tab:tnu} 
lists the values of $t_{\nu}$ for our models.

We take the $z$-axis to be perpendicular to the disc plane. In the vertical direction, the density profile is
\begin{equation}
\rho(R,z,t) = \frac{\Sigma}{\sqrt{2\pi}H} \exp \left(-\frac{z^{2}}{2H^{2}}\right),
\end{equation}
where $H$ is the vertical scale height.
We assume that the ring has a constant aspect ratio $h$ with $R$ and $t$, 
defined as $h\equiv H/R$.

Aside from the effect of viscosity, the disc is in rotational equilibrium and, after accounting for the asymmetric drift, it rotates with a velocity given by
\begin{equation}
v_{g}^{2} (R)= v_{K}^{2}\left(1-h^{2}+h^{2}\frac{d\ln\Sigma}{d\ln R}\right),    
\end{equation}
where $v_{K}^{2}=GM_{\bullet}/R$ is the Keplerian velocity.
In vectorial form, we have 
\begin{equation}
\vecv_{g}(R)=\pm v_{g} \hat{\vece}_{\phi}.
\label{eq:vphi_disc}
\end{equation}
The sign indicates the direction
of rotation. Our convention is that if the sign is positive, the ring rotates counterclockwise.

\begin{figure}
  \vspace{0pt}
  \includegraphics[angle=0,width=0.99\columnwidth]{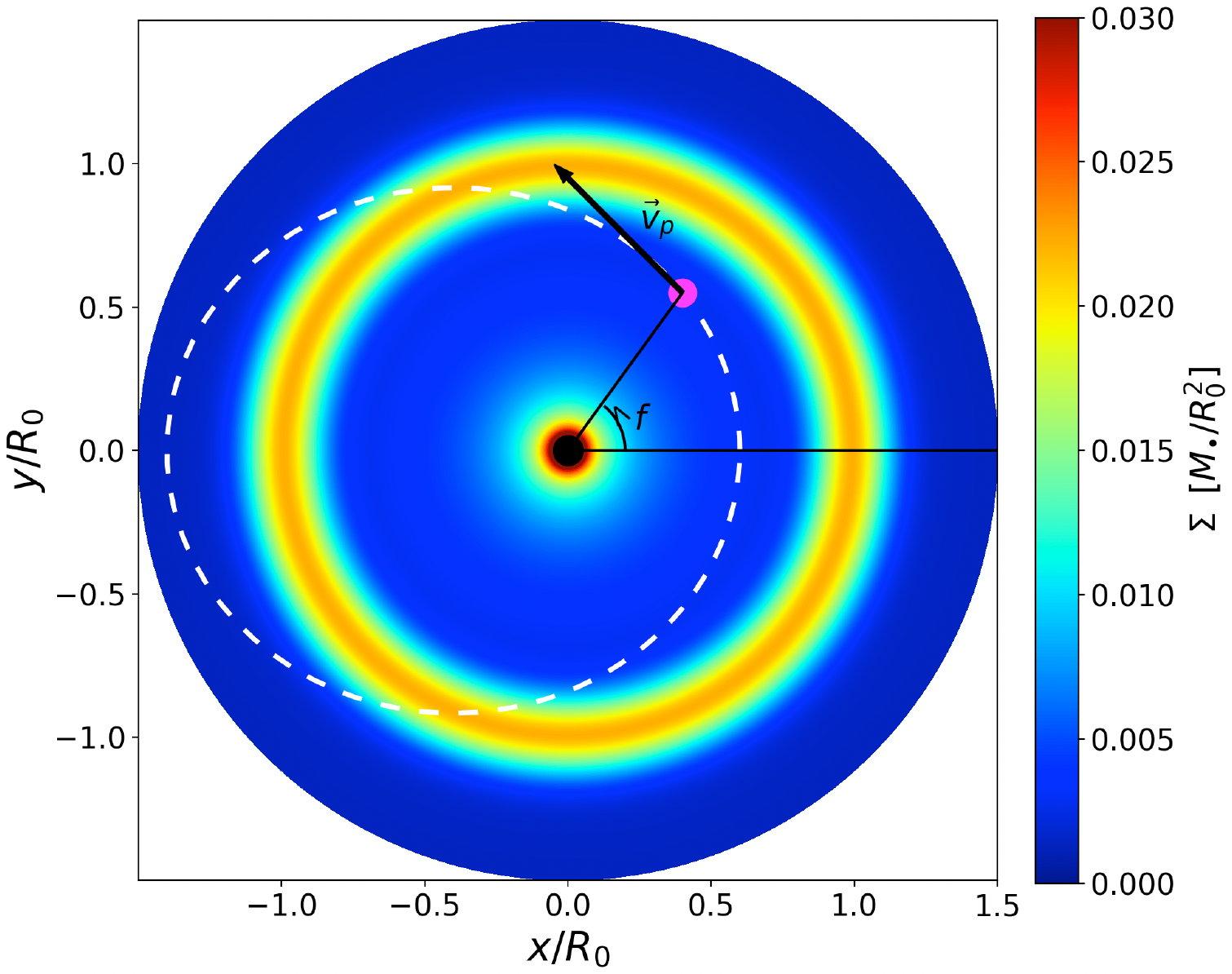}
  \caption{Surface density of the disc in model 1 at $t=0$ (colour map), together with the orbit of a perturber with $a_{p}=1$ and $e=0.4$ (white dashed ellipse), its instantaneous velocity vector $\vecv_{p}$, and the true anomaly $f$. }
  \label{fig:sketch}
\end{figure}

\subsection{The embedded perturber}
We consider a body (or perturber) of mass $M_{p}$ that moves in a Keplerian orbit around $M_{\bullet}$, with a semi-major
axis $a_{p}$ and eccentricity $e$. The perturber's orbit is assumed to be coplanar with the disc 
(the orbital plane is $z=0$)
and to proceed in a counterclockwise sense. 
The perturber's velocity in the orbital plane, expressed in polar coordinates ($\hat{\vece}_{R}$ , $\hat{\vece}_{\phi}$), is given by:
\begin{equation}
\vecv_{p} = \frac{e\omega_{p} a_{p}}{\eta}\sin f \,\hat{\vece}_{R}  + \frac{\omega_{p} a_{p}}{\eta}(1+e \cos f )
\,\hat{\vece}_{\phi},
\label{eq:vp}
\end{equation}
where $f$ is the true anomaly ($f=0$ corresponds to the pericentre), $\omega _{p}= \sqrt{GM_{\bullet}/a_{p}^{3}}$ and $\eta\equiv 
\sqrt{1-e^{2}}$. For simplicity and without loss of generality, we orient the $x$ and
$y$ axes so that the pericentre is located at $x=(1-e)a_{p}$ and $y=0$. The configuration is illustrated in Figure \ref{fig:sketch}.

From Eqs. (\ref{eq:vphi_disc}) and (\ref{eq:vp}), we see that a positive sign in Equation~(\ref{eq:vphi_disc}) indicates that the disc and the body rotate in the same direction (prograde case), while a negative sign implies counter-rotation (retrograde case).

The relative velocity of the local gas with respect to
the perturber, $\vecv_{\rm rel}\equiv \vecv_{g}-\vecv_{p}$, is supersonic (i.e. larger than
the isothermal sound speed $c_{s}$) along the entire orbit if
$e\gtrsim 2h$ for the prograde case \citep[e.g.,][]{mut11}. More specifically, in an isothermal disc,
the Mach number $\mathcal{M}\equiv v_{\rm rel}/c_{s}$ is $\gtrsim e/(2h)$ throughout the orbit. 
In the retrograde case, the motion is always supersonic,
regardless the eccentricity \citep{san20}.

Consider perturbers that move supersonically $\mathcal{M}>1$.
In order not to produce a gap in the surface density of the disc, we require that the 
accretion radius should be much smaller than $H$:
\begin{equation}
\frac{2GM_{p}}{c_{s}^{2}(1+\mathcal{M}^{2})} \ll H
\end{equation}
\citep[e.g.,][]{kru06,san23,san25}. In terms of $q\equiv M_{p}/M_{\bullet}$, the above
condition implies
\begin{equation}
q \ll \frac{1}{2} (1+\mathcal{M}^{2}) h^{3}.
\label{eq:lowmass}
\end{equation}
For a typical value of $h=0.05$ and for $\mathcal{M}\simeq 1$, the above condition
can be cast as $q\ll 10^{-4}$, whereas for the same value of $h$ but $\mathcal{M}=3$,
it requires $q\ll 6\times 10^{-4}$. As a reference value we will take $q=10^{-5}$,
which may represent a $3M_{\oplus}$ planetary core around a star of one solar mass, 
or a black hole of $10M_{\odot}$ around a supermassive black hole
of $10^{6}M_{\odot}$. For typical values of the viscosity in protoplanetary or AGN
accretion discs, a perturber with $q=10^{-5}$ cannot open a deep gap even on a 
circular orbit, whether prograde \citep[e.g.,][]{duf15} or retrograde \citep{iva15,san25}.

\subsection{Dynamical friction force and accretion force}
We assume that the object interacts with the disc solely via gravitational forces,
while feedback effects, such as thermal torques, will be examined in Section \ref{sec:applicability}.
If the object moves supersonically, it induces an overdense wake, which exerts
a gravitational force on the body \citep[e.g.,][]{san19}. In addition to this gravitational force, mass accretion
also produces a transfer of momentum from the disc gas to the perturber.
\citet{can13} find that the total drag force (gravitational plus accretion)
on a body moving hypersonically in rectilinear orbit in the midplane of a Gaussian layer is:
\begin{equation}
\vecF_{T} = \frac{2\sqrt{2\pi} \Sigma (GM_{p})^{2}}{v_{\rm rel}^{3} H}
\ln\left(\frac{3.6 H}{\xi_{0}}\right) \vecv_{\rm rel},
\label{eq:canto1}
\end{equation}
where $\xi_{0}\equiv GM_{p}/v_{\rm rel}^{2}$. In terms of the midplane density $\rho_{0}$, $h$ and $\mathcal{M}$
\begin{equation}
\vecF_{T}= 
\frac{4\pi \rho_{0} (GM_{p})^{2}}{v_{\rm rel}^{3}}
\ln\left(\frac{3.6 h^{3}\mathcal{M}^{2}}{q}\right) \vecv_{\rm rel}.
\label{eq:canto2}
\end{equation}
For fixed $v_{\rm rel}$, $F_{T}$ exhibits only a logarithmic dependence on $\mathcal{M}$.
However, when $c_{s}$ is fixed and $v_{\rm rel}$ varies, one finds $F_{T}\propto \mathcal{M}^{-2}$.

\begin{figure}
  \vspace{0pt}
  \includegraphics[angle=0,width=0.88\columnwidth]{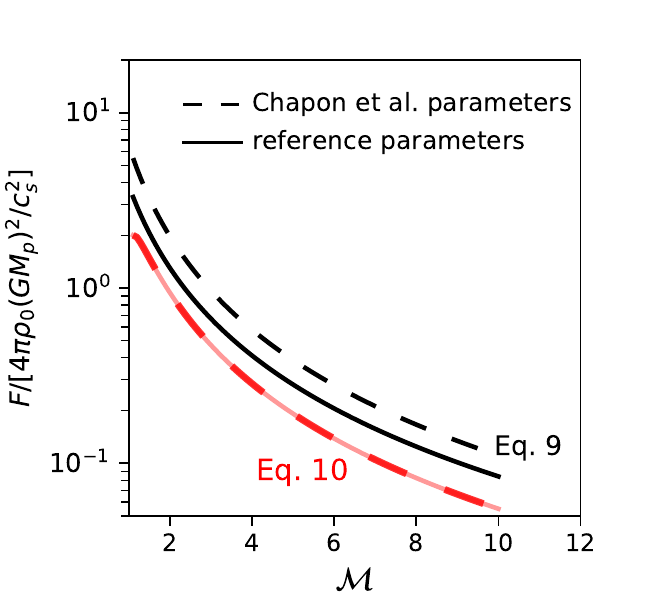}
  \caption{Comparison of the dimensionless drag force obtained with Equation (\ref{eq:canto2}) (black lines) and Equation (\ref{eq:chapon}) (red lines). Dashed lines correspond to the parameters 
  used in \citet{cha13}, namely $q=1$, $H=360$ pc and $c_{s}=75$ km s$^{-1}$. The solid lines represent the drag force calculated for the fiducial parameters adopted in the present work ($q=10^{-5}$ and $h=0.05$). Note that the two red lines overlap.}
  \label{fig:force_comparison}
  %This figure is created by mac: Documents/disco6/tex/Rodrigo_Anaya/manuscript/simple_codes/comparison_forces_v5.py
\end{figure}

\begin{figure*}
  \centering
\hspace*{-0.18\columnwidth} 
\includegraphics[scale=0.6,angle=0]{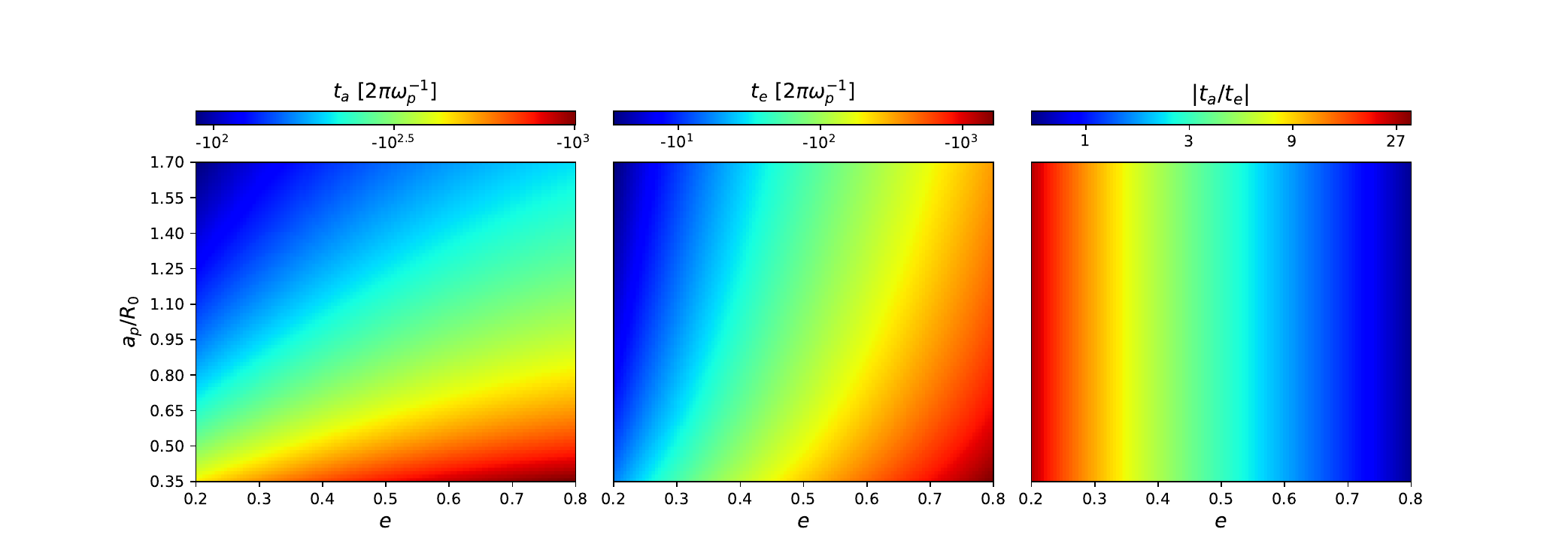}
  \caption{Colour maps of $t_{a}$ (left panel), $t_{e}$ (central panel) and the ratio $|t_{a}/t_{e}|$ (right panel) for a prograde perturber with $q=10^{-5}$ in a disc without a ring-like structure (i.e. $M_{\rm ring}=0$). The parameters
of $\beta$ and $\Sigma_{\rm bk}$ are the same as model 1 (see Table \ref{tab:parameters}).}
  \label{fig:ta_te_init_model0}
  %This figure is created by mac: Documents/disco6/tex/Rodrigo_Anaya/viscous_disk_v2/maps_ta_te/fig_ta_te_model0_v2.py
\end{figure*}

Using numerical simulations of non-accreting black hole binary embedded in a disc,
\citet{cha13} find that for $\mathcal{M}>1$, the dynamical friction force is
\begin{equation}
\vecF_{\rm CMT} = \frac{4\pi \rho_{0} (GM_{p})^{2}}{v_{\rm rel}^{3}} \left(\frac{1}{2}
\ln(\mathcal{M}^{2}-1) +3.2\right) \vecv_{\rm rel}.
\label{eq:chapon}
\end{equation}
\citet{cha13} fixed the mass of the black holes to $2.5\times 10^{6}M_{\odot}$,
the disc has a sound speed of $75$ km/s and $H=360$ pc. For these values, 
we can compute $F_{T}$ using Eq. (\ref{eq:canto2}) as a function
of Mach number and compare it with Eq. (\ref{eq:chapon}) (see Fig. \ref{fig:force_comparison}). We see that the dependence on Mach number
is similar. However, the magnitude of the forces is different; $F_{T}$ is about a factor
of $2.5$ larger than the dynamical friction force using Eq. (\ref{eq:chapon}). However,
note that \citet{cha13} 
do not include accretion in their simulations, and hence a part of the difference arises
because they ignore the contribution of accretion to the drag force.

\citet{cha13} do not investigate how the drag force depends on $M_{p}$ or $H$
(or, equivalently, on $q$ and $h$). Assuming that Eq. (\ref{eq:chapon}) also holds
for the reference values adopted in this study ($q=10^{-5}$ and $h=0.05$), 
the values predicted by Eq. (\ref{eq:canto2}) are larger than predicted by
Eq. (\ref{eq:chapon}) by a factor of $1.5$ (see Figure \ref{fig:force_comparison}).
In this work, we adopt $\vecF_{T}$ (Eq. \ref{eq:canto2}) to represent the 
drag force. We note that,  had Equation (\ref{eq:chapon}) been used instead, the characteristic timescales governing the orbital evolution of the perturbers would
increase by approximately a factor of $\sim 1.5$.

Equation (\ref{eq:canto2}) was originally derived for straight-line orbits in a medium 
of uniform surface density. In the present work, however, we apply it to bodies
on eccentric orbits embedded in
a disc where the surface density varies with $R$. For low-mass perturbers
(see condition in Eq. \ref{eq:lowmass}), this local approximation remains justified
since a significant contribution to the drag force arises from the wake within distances 
$\lesssim H$ of the perturber \citep{san18,san19}.

\subsection{Timescales for $a_{p}$ and $e$}
The evolution of the semi-major axis $a_{p}$ and eccentricity $e$ due to the interaction 
with the disc are given by
\begin{equation}
\frac{da_{p}}{dt} = \frac{2P}{\omega_{p}^{2}a_{p}},
\label{eq:da_dt1}
\end{equation}
and
\begin{equation}
\frac{de}{dt}= \frac{\eta^{2}}{\omega_{p}^{2} a_{p}^{2}e} \left(P- \frac{\omega_{p}}{\eta}T\right),
\end{equation}
where $P$ is the specific power
\begin{equation}
P= \frac{1}{M_{p}}\vecv_{p}\cdot \vecF_{T},
\end{equation}
$T$ is the magnitude of the specific torque
\begin{equation}
T=\frac{1}{M_{p}}(\vecr_{p}\times \vecF_{T})\cdot \hat{\vece}_{z},
\end{equation}
and $\eta\equiv \sqrt{1-e^{2}}$
\citep[e.g.,][]{mut11}.

If $a_{p}$, $e$ and $\Sigma$ vary slowly with time, that is,
$da_{p}/dt \ll a_{p}\omega_{p}$, $de/dt\ll e\omega_{p}$
and $d\Sigma/dt \ll \Sigma \omega_{p}$, we can compute the temporal variation of $a_{p}$ and $e$ averaged over one orbit:
\begin{equation}
\left<\frac{da_{p}}{dt}\right> \equiv \frac{1}{T_{\rm orb}}\int_{0}^{T_{\rm orb}} \frac{da_{p}}{dt}dt = \frac{1}{T_{\rm orb}}\int_{0}^{2\pi}
\frac{da_{p}}{dt}\left(\frac{df}{dt}\right)^{-1} df.
\label{eq:da_dt_1orbit}
\end{equation}
To express the equation in terms of the true anomaly $f$, we performed the
change of variable $t\rightarrow f$, 
as both $\vecr_{p}$ and $\vecv_{p}$, and hence $P$ and $T$, are functions of $f$ \citep[see also][]{mut11}. During orbital averaging, 
the surface density is assumed to depend only on radius $R$ and is treated as constant over the integration time interval.
An analogous expression holds for
$e$.

At a given $t$, we define the (instantaneous) characteristic timescales $t_{a}$ and $t_{e}$
as
\begin{equation}
t_{a}\equiv \frac{a_{p}}{\left<da_{p}/dt\right>},
\end{equation}
and
\begin{equation}
t_{e}\equiv \frac{e}{\left<de/dt\right>}.
\end{equation}
A negative (positive) timescale indicates that the quantity
is decreasing (increasing) with time.

Given the initial semi-major axis and eccentricity, along with the time-dependent surface density (Eq. \ref{eq:Sigma}),
Equation (\ref{eq:da_dt_1orbit}) and the
counterpart for the eccentricity can be integrated to obtain $a_{p}(t)$ and $e(t)$.

\section{Results}
\label{sec:results}
We will consider two very different situations: when the perturber rotates in the same
direction as the disc, and the retrograde case, where the perturber rotates in the opposite sense. 
We discuss these cases in two separate subsections. Unless stated otherwise, we adopt
the fiducial values $q=10^{-5}$ and $h=0.05$.

\subsection{Prograde orbits}

\begin{figure}[
  \vspace{0pt}
  \includegraphics[angle=0,width=0.99\columnwidth]{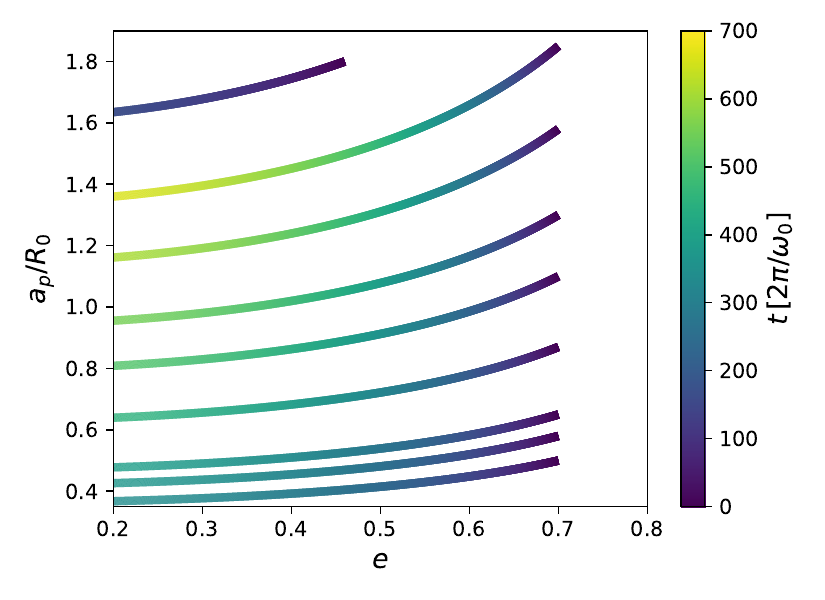}
  \caption{Trajectories in the $(e,a_{p})$ plane for the same model as in Figure \ref{fig:ta_te_init_model0}, i.e. the disc has a power-law density, $\Sigma= \Sigma_{\rm bk} (R_{0}/R)$, with $\Sigma_{\rm bk}=2\times 10^{-3}M_{\bullet}/R_{0}^{2}$, as in
model 1, but with $M_{\rm ring}=0$. The perturber has $q=10^{-5}$.}
  \label{fig:evol_model0}
\end{figure}

For prograde orbits, we restrict our analysis to orbital eccentricities greater than $0.2$ to ensure
that $\mathcal{M}$ exceeds $2$ at all orbital positions (for our fiducial value $h=0.05$).
To better highlight the impact of the ringed structure, it is instructive to first examine
the timescales $t_{a}$ and $t_{e}$ for a disc without any ring component.
Figure \ref{fig:ta_te_init_model0} presents these timescales for $\beta=1$ and
$\Sigma_{\rm bg}= 2\times 10^{-3}M_{\bullet}/R_{0}^{2}$, as in model 1, but
with $M_{\rm ring}=0$. In this case, both $t_{a}$ and $t_{e}$ are negative,
indicating inward migration and continuous eccentricity damping. For a given $e$,
the two timescales vary as $a_{p}^{-1}$. The corresponding temporal evolution
of $a_{p}$ and $e$ is illustrated in Figure \ref{fig:evol_model0}.

\begin{figure*}
  \centering
\hspace*{0.0\columnwidth} 
\includegraphics[trim=150 0 160 0, clip, scale=0.64,angle=-90]{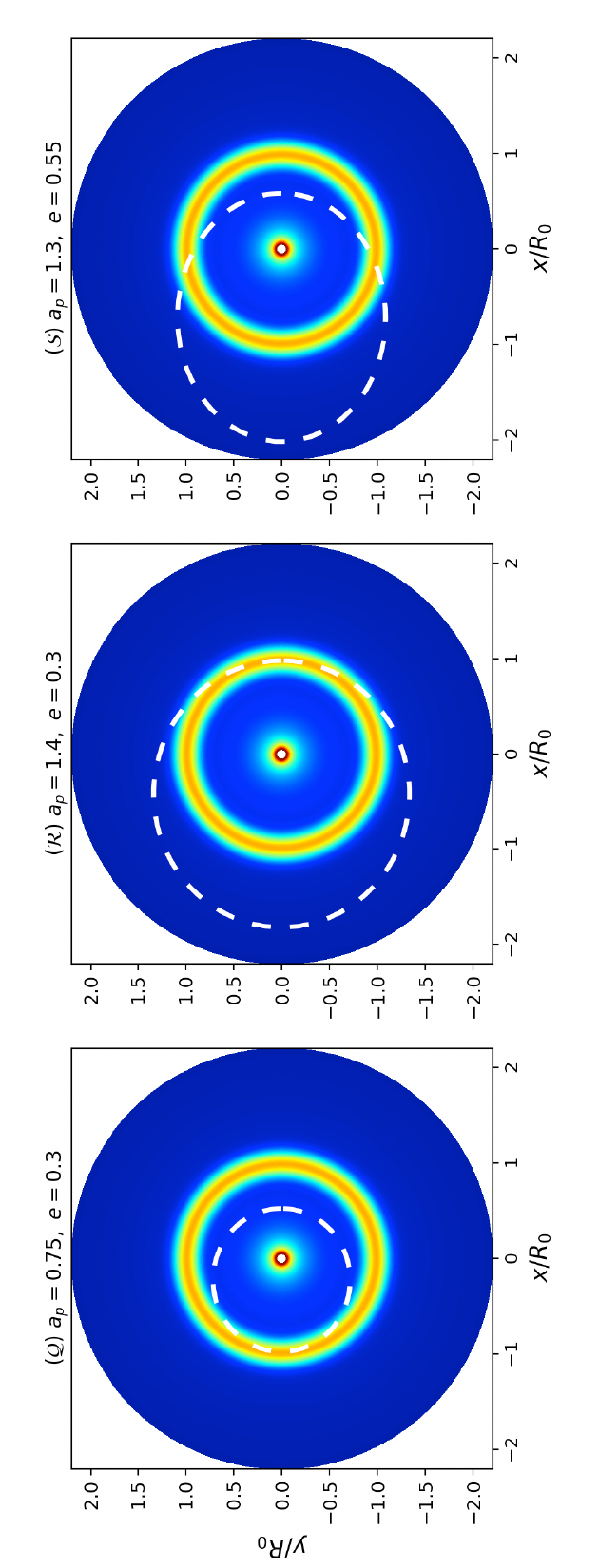}
  \caption{Three representative orbits are shown. The left panel corersponds to the case labeled $\mathcal{Q}$, for brevity. In this configuration, the apocentre lies near the ring's maximum. In the central panel (case labeled $\mathcal{R}$), the pericentre coincides with the ring. In the right panel (case $\mathcal{S}$), the perturber's orbit crosses the ring.}
  \label{fig:3orbits}
  %This figure is created by mac: Documents/disco6/tex/Rodrigo_Anaya/viscous_disk_v2/sketch/sketch_3orbits.py
\end{figure*}

\begin{figure*}
  \centering
\hspace*{-0.\columnwidth} 
\includegraphics[scale=0.58,angle=0]{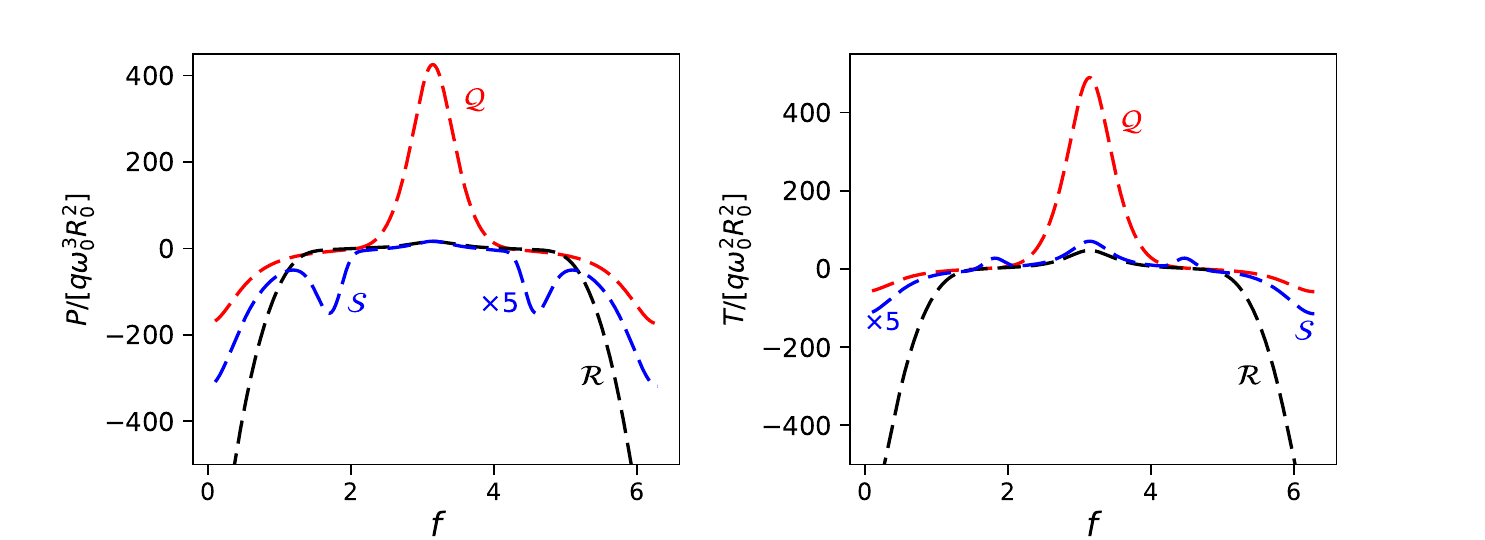}
  \caption{Specific power (left panel) and torque (right panel) versus true anomaly, for case 
$\mathcal{Q}$ ($a_{p}=0.75$, $e=0.3$; red lines),
case $\mathcal{R}$ ($a_{p}=0.75$, $e=0.3$; blue lines) and case $\mathcal{S}$ ($a_{p}=1.3$, $e=0.55$; black lines) for prograde orbits in model 1 at $t=0$. In case $\mathcal{S}$, the curves have been scaled by a factor of $5$ to enhance visibility.}
  \label{fig:pw_tq_model1}
  %This figure is created by mac: Documents/disco6/tex/Rodrigo_Anaya/viscous_disk_v2/maps_ta_te/fig_power_tq_model1.py
%The data is obtained by running Documents/disco6/tex/Rodrigo_Anaya/viscous_disk_v2/maps_ta_te/pw_tq_model1.py
\end{figure*}

We now turn to the case of a disc featuring a ring structure.
Figure \ref{fig:3orbits} shows three representative orbits that differ in the relative positions
of their pericentres and apocentres with respect to the ring. For brevity,
we label these orbits as $\mathcal{Q}$, $\mathcal{R}$ and $\mathcal{S}$, 
respectively.
Figure \ref{fig:pw_tq_model1} shows the power $P$ and the torque $T$ as a function of $f$ in model 1 at $t=0$, assuming prograde orbits. In case $\mathcal{Q}$, where
the perturber's orbit is tangential to the ring at apocentre ($f=\pi$),  both $P$ and $T$ are large and positive there. In contrast, in case $\mathcal{R}$, where the orbit is tangential
at pericentre ($f=0$), both $P$ and $T$ are large and negative there. In case $\mathcal{S}$,  the power exhibits valleys at the phase angles were the perturber crosses the ring ($f\simeq 1.8$ and $f\simeq 4.5$). However, the power amplitude during ring crossing is about an order of magnitude smaller than in cases $\mathcal{Q}$ and $\mathcal{R}$.
Similarly, the magnitude of the torque in case $\mathcal{S}$ during the ring crossing is 
significantly lower than in cases $\mathcal{Q}$ and $\mathcal{R}$. The reason for this suppresion in the power and torque is that the
perturber passes through the ring with a predominantly radial relative velocity. 
These differences between orbits $\mathcal{Q}$, $\mathcal{R}$ and $\mathcal{S}$ 
indicate that the evolution of $a_{p}$ and $e$ depends on the relative position of pericentres and apocentres with respect to the ring.

\begin{figure*}
  \centering
\hspace*{-0.18\columnwidth} 
\includegraphics[scale=0.6,angle=0]{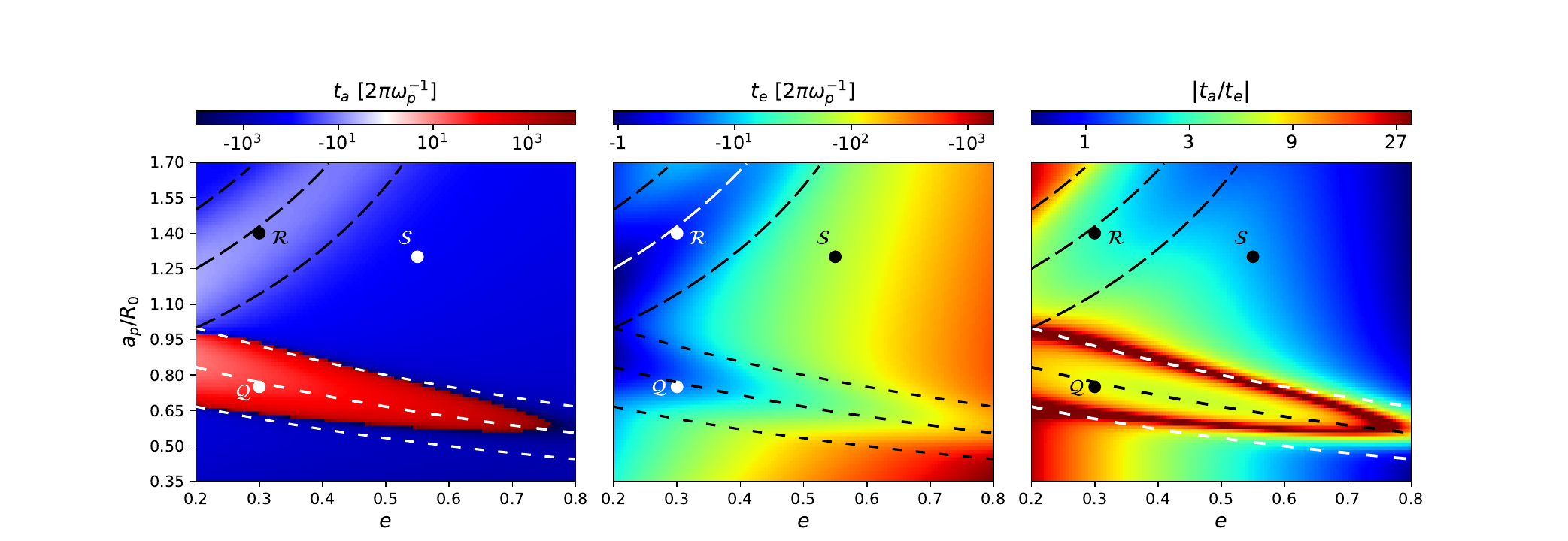}
  \caption{Colour maps of $t_{a}$ (left panel), $t_{e}$ (central panel) and the ratio $|t_{a}/t_{e}|$ (right panel) for a prograde perturber with $q=10^{-5}$ in model 1 at $t=0$. The three upper
  ascending curves indicate those orbits with pericentres at $1.2R_{0}, R_{0}$ and $0.8R_{0}$ (from top to bottom), whereas the three lower curves correspond to orbits with apocentres
  at $1.2R_{0}$, $R_{0}$ and $0.8R_{0}$ (from top to bottom). The different colours of these curves are assigned purely for readability and have no physical significance. The position of the orbits  
$\mathcal{Q}$, $\mathcal{R}$ and $\mathcal{S}$ are marked by dots.}
  \label{fig:ta_te_init_model1}
  %This figure is created by mac: Documents/disco6/tex/Rodrigo_Anaya/viscous_disk_v2/maps_ta_te/fig_ta_te_model1_v10.py
\end{figure*}

Figure \ref{fig:ta_te_init_model1} shows $t_{a}$ and $t_{e}$, again for model 1 at $t=0$,
as a function of $a_{p}$ and $e$. A comparison with Figure \ref{fig:ta_te_init_model0}
reveals that the ring has a strong impact on $t_{a}$,
but note that the colour scale covers a different range.
We observe an elongated region where $t_{a}>0$ (outward migration).
We notice that most of the orbits with apocentres between $0.9R_{0}$ and $1.2R_{0}$ (as in case $\mathcal{Q}$), except if they have large eccentricities, fall within this region. Outside this region,
$t_{a}<0$ (inward migration). The inward migration rate is the fastest when the pericentre
occurs at $R=R_{0}$. 

On the other hand, $t_{e}$ is always negative implying that the 
eccentricity is always damped. We see that the blue regions in the $t_{e}$ map are
elongated along the curves delineated by 
the orbits with apocentres or pericentres $\sim R_{0}$, indicating that $|t_{e}|$ is the shortest along these orbits.

\begin{figure*}
  \vspace{0pt}
\hspace*{-0.03\columnwidth} 
  \includegraphics[angle=0,width=1.0\textwidth]{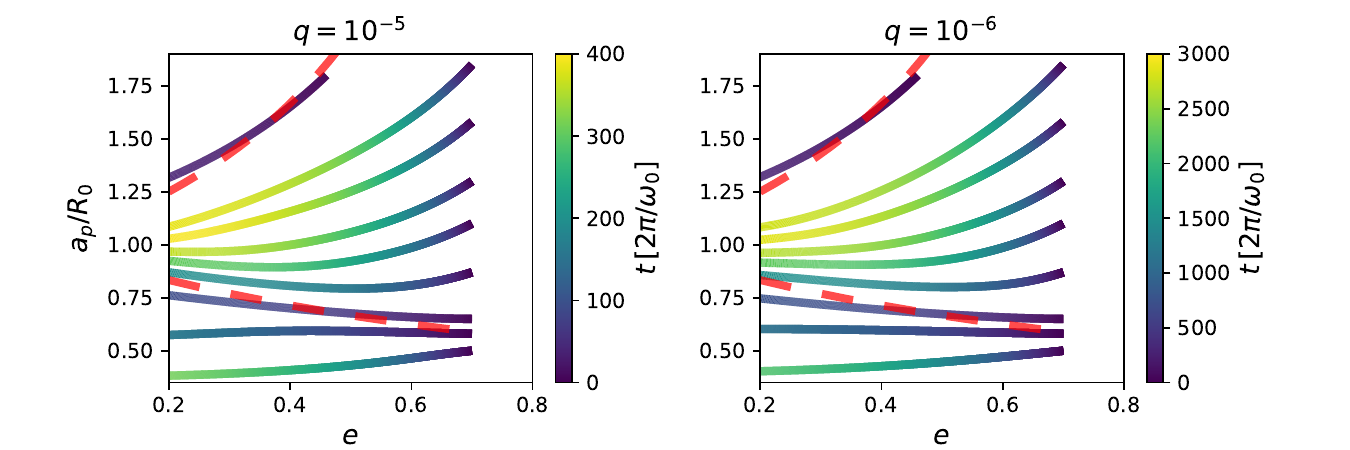}
  \caption{Trajectories of prograde orbits in the parameter plane $(e, a_{p})$ in model 1B,
for $q=10^{-5}$ (left panel) and $q=10^{-6}$ (right panel). 
The upper  and lower red dashed lines mark the orbits whose 
pericentres and apocentres, respectively, lie at $R=R_{0}$ (the ring radius).}
  \label{fig:evol_a_e_model1B_diff_q}
 %This figure is created by mac: Documents/disco6/tex/Rodrigo_Anaya/viscous_disk_v2/
%evolution_a_e/fig_evolution_v4.py. The data is created by changing initial parameters in
%viscous_ring_power_tq_adot_edot_evol_model1_v5.py
\end{figure*}

In the third panel of Fig. \ref{fig:ta_te_init_model1} we see that, except for $e>0.65$, 
$|t_{e}|\ll |t_{a}|$. In these cases, the orbit tends to circularize faster than the semi-major
axis changes. In particular, for case $\mathcal{Q}$, $|t_{a}|\simeq 8 |t_{e}|$, whereas for case $\mathcal{R}$, 
$|t_{a}|\simeq 3|t_{e}|$. As a result, the orbit in case $\mathcal{Q}$ will tend to adopt a circular orbit
within the ring, i.e. $a_{p}\leq R_{0}$, while orbit $\mathcal{R}$ will tend to a circular orbit outside
the ring, i.e. $a_{p}\geq R_{0}$. This is illustrated in Figure \ref{fig:evol_a_e_model1B_diff_q}, where
we show the trajectories of the orbits in the $(e,a_{p})$ plane in model 1B.
For $q=10^{-5}$ (left panel of the figure), the eccentricity is damped to $0.2$
in a timescale $\lesssim 400 (2\pi/\omega_{0})$. During this interval, the ring has barely
spread radially, as $t_{\nu}\simeq 3\times 10^{3} (2\pi/\omega_{0})$
(see Table \ref{tab:tnu}). As a result of this large spreading time, the geometry of the trajectories in 
the $(e,a_{p})$ plane remain unchanged: 
any orbit starting from a point along a given curve will continue to trace that
same curve.

\begin{figure*}
  \centering
\hspace*{-0.1\columnwidth} 
\includegraphics[scale=0.55,angle=0]{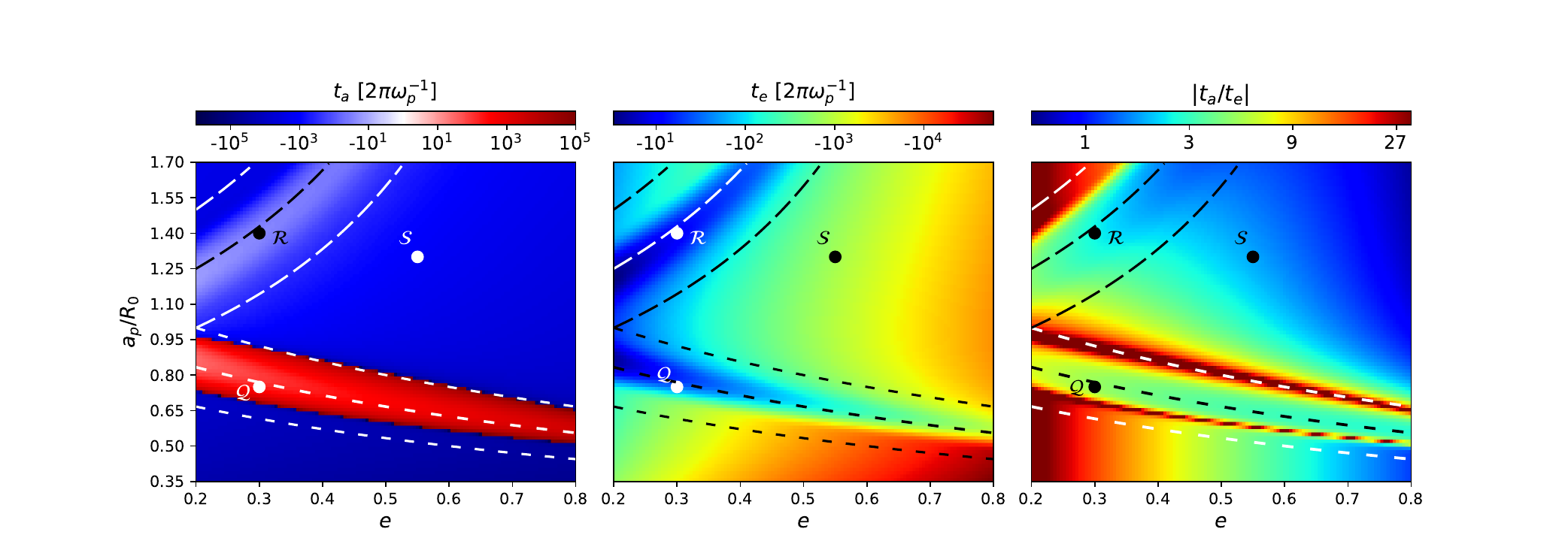}
  \caption{Similar to Figure \ref{fig:ta_te_init_model1}, but for model 2 (at $t=0$). The orbits are assumed to be prograde.}
  \label{fig:ta_te_init_model2}
  %This figure is created by mac: Documents/disco6/tex/Rodrigo_Anaya/viscous_disk_v2/maps_ta_te/fig_ta_te_model2_v8.py
\end{figure*}

The orbits located between the two red dashed lines in Figure \ref{fig:evol_a_e_model1B_diff_q} correspond 
to those orbits crossing the ring maximum. As time progresses, the semi-major axes of all such orbits tend to approach to $R_{0}$. In other words, 
the orbits become more circular and since $a_{p}\rightarrow R_{0}$, they
more closely follow the ring. Specifically, orbits that start with $a_{p}$ between $0.65R_{0}$ and $1.8R_{0}$ at $e=0.7$ evolve to
$a_{p}$ values between $0.75R_{0}$ and $1.1R_{0}$ once their eccentricities decay to $e=0.2$. Hence, 
perturbers tend to accumulate, forming an overdense, wide ring centred around $R=R_{0}$.
Our model cannot be applied to determine the subsequent evolution once the eccentricities drop
below $0.2$. If only resonant torques (i.e. Lindblad and corotation torques)
are taken into account, perturbers on circular orbits migrate
toward the trapping radius, located slightly interior to $R=R_{0}$ \citep{mas06,rom19,cha24}.
Accordingly, we expect the perturbers forming the wide ring will 
eventually migrate toward the trapping radius once their eccentricities are damped below $0.2$. 
However, in the case of low-mass planets, thermal torques could alter this conclusion 
\citep{chr23}.

The right panel in Figure \ref{fig:evol_a_e_model1B_diff_q} shows the evolution curves for $q=10^{-6}$.
We see that they are very similar to those for $q=10^{-5}$, 
differing only in that the curves for $q=10^{-6}$ proceed at
a slower rate. For these values of $q$, models 1A evolve indistinguishably from
models 1B and are thus omitted.

\begin{figure*}
  \vspace{0pt}
\hspace*{-0.03\columnwidth} 
  \includegraphics[angle=0,width=1.0\textwidth]{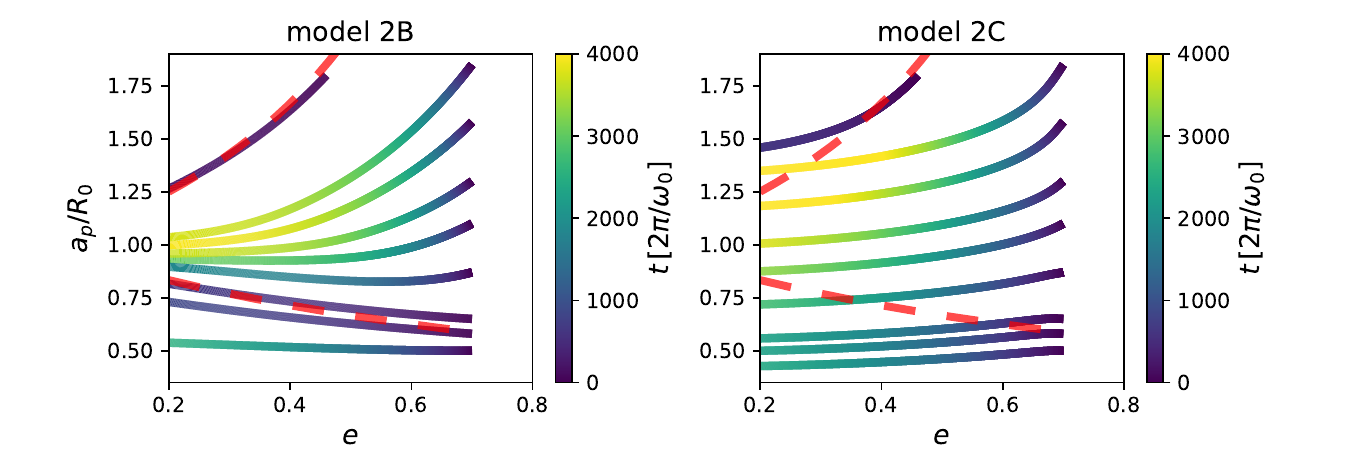}
  \caption{Trajectories of prograde orbits in the parameter plane $(e, a_{p})$ in model 2B
(left panel) and model 2C (right panel). In both cases we assume $q=10^{-5}$. 
The red dashed lines have the same meaning as in Figure \ref{fig:evol_a_e_model1B_diff_q}.}
  \label{fig:evol_a_e_model2B_2C}
 %This figure is created by mac: Documents/disco6/tex/Rodrigo_Anaya/viscous_disk_v2/
%evolution_a_e/fig_evolution_v4.py. The data is created by changing initial parameters in
%viscous_ring_power_tq_adot_edot_evol_model1_v5.py
\end{figure*}

\begin{figure*}
  \centering
\hspace*{-0.\columnwidth} 
\includegraphics[scale=0.68,angle=0]{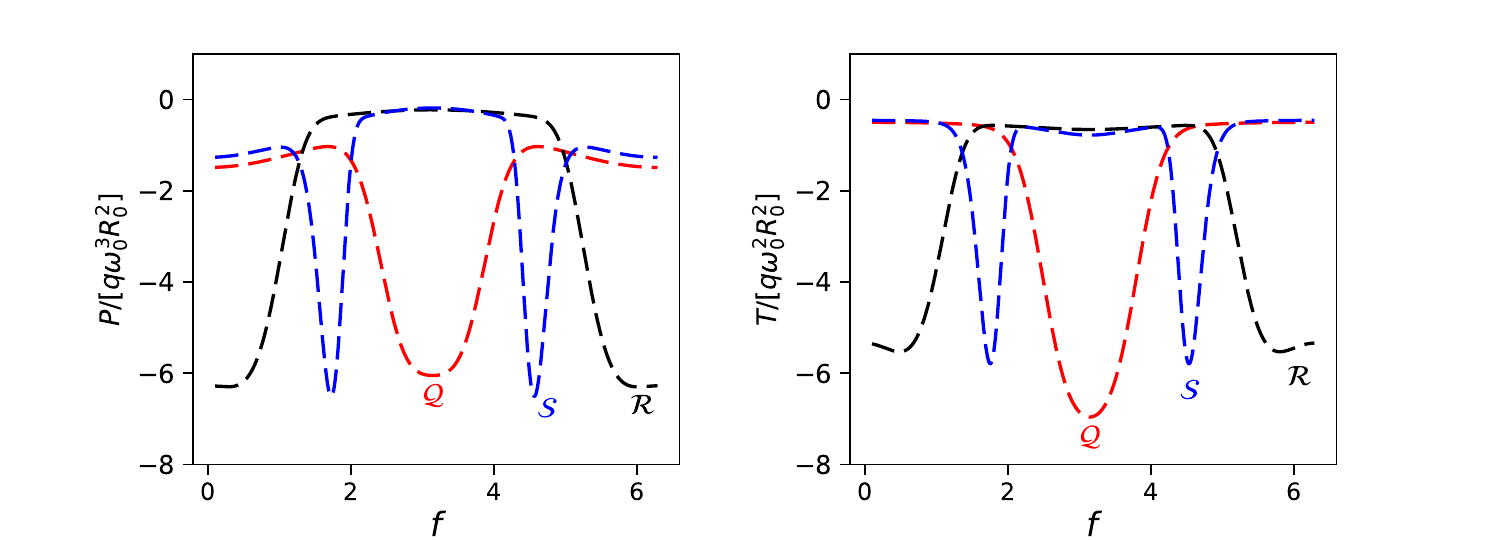}
  \caption{Specific power (left panel) and torque (right panel) versus true anomaly, similar to
Figure \ref{fig:pw_tq_model1}, but now for retrograde orbits. We used model 1 at $t=0$.}
  \label{fig:pw_tq_model1_retrograde}
  %This figure is created by mac: Documents/disco6/tex/Rodrigo_Anaya/viscous_disk_v2/maps_ta_te/fig_power_tq_model1.py
%The data is obtained by running Documents/disco6/tex/Rodrigo_Anaya/viscous_disk_v2/maps_ta_te/pw_tq_model1.py
\end{figure*}  

\begin{figure*}
  \centering
\hspace*{-0.1\columnwidth} 
\includegraphics[scale=0.55,angle=0]{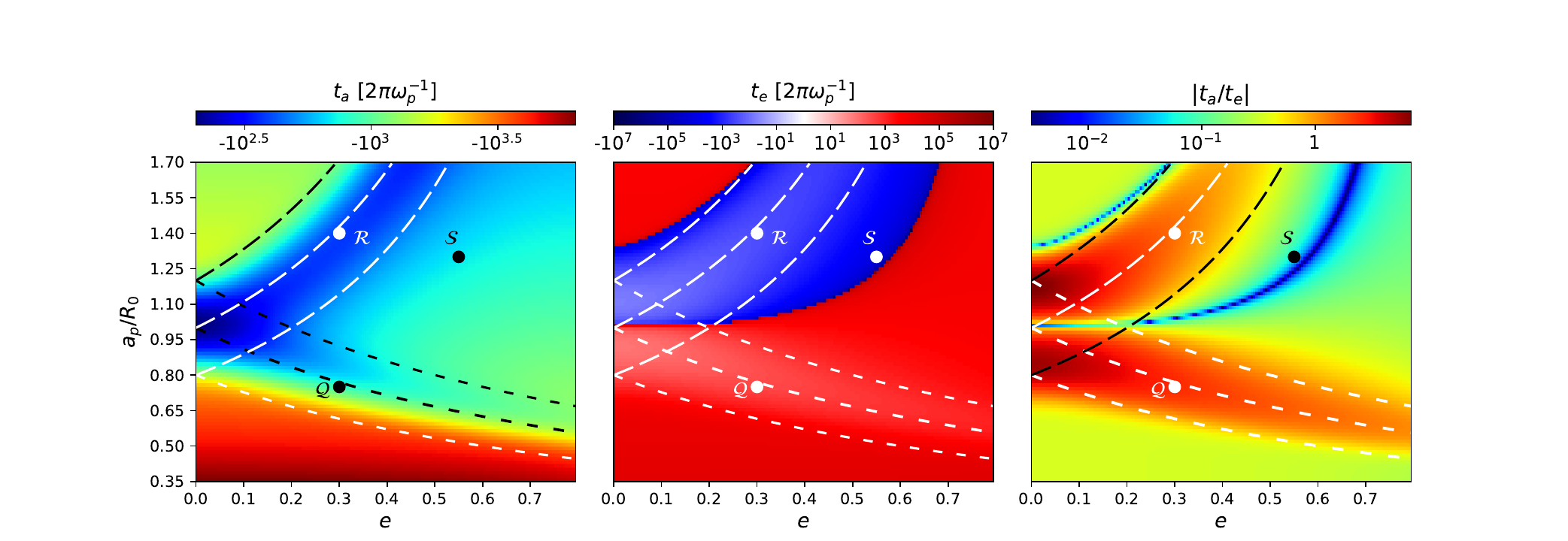}
  \caption{Similar to Figure \ref{fig:ta_te_init_model1} (model 1 at $t=0$ and $q=10^{-5}$) but assuming that the orbits are retrograde. }
  \label{fig:ta_te_init_model1_retrograde}
  %This figure is created by mac: Documents/disco6/tex/Rodrigo_Anaya/viscous_disk_v2/maps_ta_te/fig_ta_te_model1_retrograde_v2.py
% The data is created by maps_ta_te/viscous_ring_power_tq_adot_edot_a08_model1_retrograde_v4.py
\end{figure*}

Interestingly, orbits that begin with either their pericentres or apocentres
at $R=R_{0}$ evolve while approximately preserving this condition. 
The same behaviour is observed for $q=10^{-6}$ (see right panel in Fig. \ref{fig:evol_a_e_model1B_diff_q}). Perturbers trapped in these orbits experience 
maximal mass accretion, as 
$\dot{M}_{p}\propto \rho_{0} v_{\rm rel}^{-3}$ \citep{hoy39,bon44}, with $\rho_{0}$
peaking in the ring and $v_{\rm rel}$ minimized at apocentre and pericentre
 \citep[e.g., see Fig 5 in][]{mut11}. 

The gas ring has a significant effect only if it does not disperse on a timescale
shorter than ${\rm min}\{t_{a},t_{e}\}$. In model 1C, the spreading time $t_{\nu}$ is
only $15.2 (2\pi/\omega_{0})$ (see Table \ref{tab:tnu}). Hence, for $q=10^{-5}$,
the spreading time can fall below ${\rm min}\{t_{a},t_{e}\}$ for certain
$(e,a_{p})$ combinations. Therefore, those orbits remain essentially unaffected
by the ring.

Figure \ref{fig:ta_te_init_model2} shows $t_{a}$ and $t_{e}$ but for model 2 at $t=0$.
We observe the same features in the maps as in model 1. In model 2, 
the red region in the left panel
and the blue regions in the central panel are more extended and slightly thinner.
Figure \ref{fig:evol_a_e_model2B_2C} displays the orbital evolution in models 2B and
2C, for $q=10^{-5}$. The trajectories in the ($e,a_{p}$) plane in model 2B resemble 
those of model 1B,
but the convergence of $a_{p}$ toward $R_{0}$ is more pronounced in model 2B.
Finally, in model 2C, the ring spreads so rapidly compared to $t_{a}$ and $t_{e}$
that the evolution of $a_{p}$ and $e$ closely resembles the case without a ring
(see Figure \ref{fig:evol_model0} for the evolution in model 1 without a ring).

To summarize, in models where $t_{\nu}$ is sufficiently
larger than both $t_{a}$ and $t_{e}$ (models 1A, 1B, 2A and 2B),  orbits that
cross the density maximum of the ring tend to circularize, with $a_{p}$ 
converging toward $R_{0}$. As a result,
perturbers accumulate and form a secondary ring superimposed to the gaseous one. 
Moreover, orbits with apocentre or pericentre at $R=R_{0}$ maintain this condition
during their evolution. By contrast, in models 1C and 2C, the ring dissolves rapidly 
and has only a minor effect on the orbital dynamics.

\subsection{Retrograde orbits}
In this section we focus on the evolution of orbits counter-rotating with respect to the disc.
Figure \ref{fig:pw_tq_model1_retrograde} shows the power and the torque versus $f$,
for orbits $\mathcal{Q}$, $\mathcal{R}$, and $\mathcal{S}$ in model 1 at $t=0$. A comparison between Figures \ref{fig:pw_tq_model1} and \ref{fig:pw_tq_model1_retrograde} indicates that the magnitudes of both the power 
and the torque
are lower in the retrograde case. This is because  the Mach number of the perturber is significantly higher in the retrograde configuration, resulting in a reduced dynamical 
friction force. The range of values taken by $P$ and $T$ over $f$
are very similar in all three cases.
In contrast to the prograde case, both the power and the torque are negative for all
values of $f$. Because $\left<P\right>$ is negative, the migration proceeds inward
(see Eq. \ref{eq:da_dt1}).

Figure \ref{fig:ta_te_init_model1_retrograde} shows $t_{a}$ and $t_{e}$, again for model
1 at $t=0$ in the retrograde case for $e>0$. 
As anticipated, $t_{a}$ is negative for any combination
of $a_{p}$ and $e$. We clearly identify two blue branches in the $t_{a}$ map. The most
prominent branch corresponds to orbits with pericentres between $0.85R_{0}$ and $R_{0}$, where the radial migration rate is high. For orbits with pericentres $\sim R_{0}$, $t_{a}$
varies between $-250$ and $-450$ perturber's orbital periods, 
depending on the value of $e$. The second branch extends
over the region of orbits with apocentres between $R_{0}$ and $1.2R_{0}$.

\begin{figure*}
  \vspace{0pt}
\hspace*{-0.0\columnwidth} 
  \includegraphics[angle=0,width=1.0\textwidth]{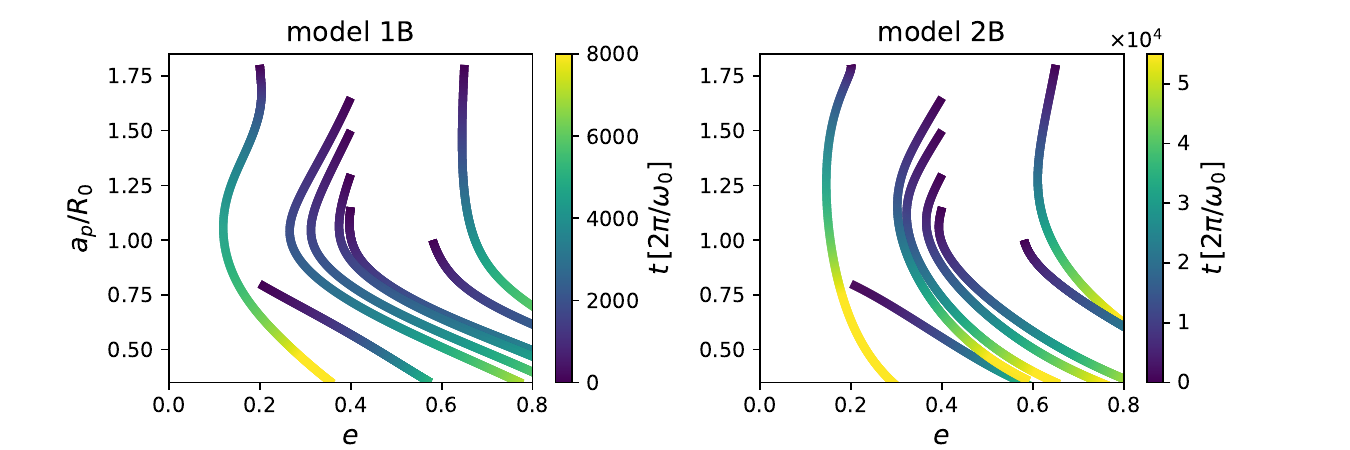}
  \caption{Trajectories of retrograde orbits in the $(e, a_{p})$ plane in model 1B
(left panel) and model 2B (right panel). In all cases we assume $q=10^{-5}$. 
}
  \label{fig:evol_a_e_model1B_2B_retrograde}
 %This figure is created by mac: Documents/disco6/tex/Rodrigo_Anaya/viscous_disk_v2/
%evolution_a_e_Sigma_retrograde/fig_model1B_2B_v2.py. The data is created by changing initial parameters in
%evol_model1B.py and evol_model2B.py
\end{figure*}

In contrast to prograde orbits, the perturber's eccentricity can be excited for some
combinations of $a_{p}$ and $e$ (red regions in the $t_{e}$ panel of Figure \ref{fig:ta_te_init_model1_retrograde}). \citet{san20} already noticed that in power-law
discs, the eccentricity may be damped or excited depending on the disc parameters.
Remarkably, the lower boundary separating the regions with $t_{e}>0$ and $t_{e}<0$,
does not coincide with orbits of constant pericentre.
For $e=0$, there exists a range of $a_{p}/R_{0}$ values between $1$ and $1.3$
where $t_{e}<0$. This does not imply that the eccentricity becomes negative. Since
$de/dt=e/t_{e}$, it follows that $de/dt\rightarrow 0$ as $e\rightarrow 0$.

In the third panel of Figure \ref{fig:ta_te_init_model1_retrograde}, we see that for most
of the combinations of orbital parameters, $|t_{a}/t_{e}|$ ranges between $0.1$ and $3$.
In particular, in case $\mathcal{Q}$, the eccentricity increases more rapidly than the 
perturber migrates inward, so the orbit moves primarily to the right and only slightly downward in the $(e,a_{p})$ plane. 
In case $\mathcal{R}$, by contrast, the eccentricity is damped at a rate comparable to 
the decay of $a_{p}$; the orbit therefore moves leftward and downward, entering the
region where eccentricity damping dominates over migration. In case $\mathcal{S}$, the radial migration proceeds faster than the change in eccentricity,
causing the orbit in the $(e, a_{p})$ plane to move downward into the red region, where
the eccentricity is instead excited.

The trajectories in the $(e,a_{p})$ plane for models
1B and 2B are shown in Figure \ref{fig:evol_a_e_model1B_2B_retrograde}. 
The orbital evolution in model 1A is essentially
indistinguishable from that in model 1B and is therefore not shown.
We see that all perturbers drift toward the central object. The eccentricity
may be damped or slightly excited when $a_{p}>R_{0}$, 
whereas for $a_{p}<R_{0}$, it invariably grows.  
In the cases shown in Figure \ref{fig:evol_a_e_model1B_2B_retrograde}, the eccentricity 
at $a_{p}=0.35 R_{0}$ is always greater than its initial value. However, it is still insufficient
for the orbits to reach the ring; that is, the orbits lie entirely inside it.

\begin{figure}
  \vspace{0pt}
  \includegraphics[angle=0,width=0.9\columnwidth]{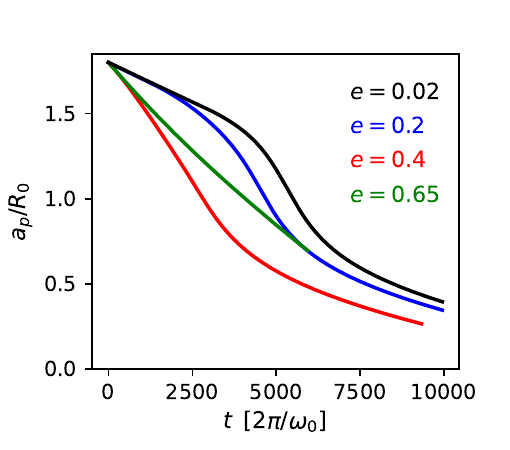}
  \caption{Temporal evolution of the semi-major axis for different values of the initial
eccentricity, for retrograde orbits in model 1B with $q=10^{-5}$. 
The curves start with $a_{p}=1.8R_{0}$ and terminate at the time when $e=0.8$.}
  \label{fig:a_t_model1B_retrograde}
%This figure is created by mac: Documents/disco6/tex/Rodrigo_Anaya/viscous_disk_v2/evolution_a_e_Sigma_retrograde/fig_a_t_v2.py
\end{figure}

Figure \ref{fig:a_t_model1B_retrograde} shows how radial migration depends on
the initial eccentricity. In all cases, we set $q=10^{-5}$ and 
$a_{p}=1.8R_{0}$ at $t=0$. Initially, the radial migration is slower at low
eccentricities, because the perturbers remain in a low-density region and do not cross 
the ring.
The fastest radial migration occurs for $e=0.4$. In this case, the perturber
reaches $a_{p}=R_{0}$ in half the time required for a perturber with an
initial eccentricity of $e=0.02$.

Finally, as in the prograde case, the ring in models 1C and 2C exerts no 
appreciable influence on the orbits, since the orbital evolution timescales are far 
longer than $t_{\nu}$.

\section{Applicability to specific astrophysical systems}
\label{sec:applicability}

The results  of the previous section rely on a series of simplistic assumptions, which
are intended as a starting point for more detailed future studies. In particular,
the perturber is modelled as a non-luminous, perfectly accreting point mass
that interacts with a smooth, unmagnetized medium solely through 
gravitational forces. In this section, we outline the possible role of additional physical
processes in shaping the drag force in two specific systems: compact objects embedded in AGN accretion discs and planets (or planetary embryos) in protoplanetary discs.

\subsection{Compact objects in AGN accretion discs }
We have assumed that the disc is smooth. However, AGN accretion discs may exhibit
some degree of clumpiness, as they are likely close to the threshold of  gravitational instability. Such clumpiness can lead to stochastic gravitational interactions, potentially
scattering the embedded compact objects and driving them onto inclined orbits \citep{sou17}.

Our assumption that compact objects (such as black holes) behave as non-luminous point masses
is widely adopted in studies of their migration within AGN accretion discs 
\citep[e.g.,][]{koc11,2016ApJ...819L..17B,2018ApJ...866...66M,2019ApJ...878...85S,2020ApJ...899..126S,2025MNRAS.544.2024J}. 
However, radiative feedback may alter the gaseous drag experienced by a 
black hole. \citet{2024MNRAS.530.2114G} examine the importance of thermal torques on black holes in circular orbits
within the radiative diffusion approximation.
In a different regime, for supersonic black holes embedded in a medium of gas number density $n_{0}$, 
radiation-hydrodynamic simulations show that radiative feedback can modify the dynamical
friction force by photoionizing and heating the gas in its vicinity, provided that 
$(1+\mathcal{M}^{2})n_{0} M_{\rm BH}<10^{9} M_{\odot}$cm$^{-3}$,
and $\mathcal{M}<4$ (\citealt{2017ApJ...838..103P} and references
therein; \citealt{2020MNRAS.496.1909T}; \citealt{2024MNRAS.528.2588O}). In the context of  
AGN accretion discs, which are very dense,
radiative feedback is expected to become inefficient due to rapid cooling and 
strong confinement of the ionized region, with the resulting wake resembling
that of the classical BHL scenario.

It has been suggested that outflows and jets launched by compact objects could reduce
the drag force they experience by partially disrupting the
overdense wake that forms behind them \citep{gru20,li20,car25}. 
\citet{li20} simulated bodies moving through a homogeneous medium while launching accretion-powered jets 
with outflow velocity $v_{w}$.
They find that, when the ratio of ejected mass rate to accreted mass rate is $0.3$, 
the drag force is reduced by up to a factor of $\simeq 0.4$, at
$u\equiv v_{p}/v_{w}=0.03$, provided the jets are launched perpendicular to the direction of motion.
In their simulations, the wake extends to $15R_{\rm acc}$, where we recall that 
$R_{\rm acc}\equiv 2GM_{p}/(c_{s}^{2}+v_{\rm rel}^{2})$. 
However, a correction for the finite computational domain is necessary. 
For a stellar-mass black hole embedded in an AGN accretion disc, the gravitational wake 
is expected to extend well
beyond the computational domain considered in \citet{li20}, potentially affecting the inferred drag force.
To estimate the corrected reduction of the total drag force, we proceed as follows.
Guided by \citet{li20}, we assume that the contribution to the drag arising from the wake
within $15R_{\rm acc}$ is reduced to $0.4$ of its value in the absence of outflows. Beyond this distance, 
we assume that the jet no longer significantly affects the gas distribution, so the outer gravitational 
wake contributes its full, unmodified drag.

\citet{li20} adopt $M_{p}=1M_{\odot}$, $v_{\rm rel}=300$ km s$^{-1}$, which
implies $R_{\rm acc}=3\times 10^{11}$ cm$=10^{-7}$ pc, much smaller than $H$ in the 
AGN accretion disc. In the absence of outflows, the contribution to 
the drag force from the wake within a radius $R_{\rm max}\equiv 15R_{\rm acc}\ll H$, 
plus the accretion drag, is
\begin{equation}
F_{\rm no-jet} (<R_{\rm max}) = \frac{4\pi \rho_{0} (GM_{p})^{2}}{v_{\rm rel}^{2}}
\ln \left(\frac{R_{\rm max}}{b_{\rm min}}\right),
\label{eq:std}
\end{equation}
with $b_{\rm min}\simeq \sqrt{e} R_{\rm acc}/4$ \citep{can11}. 
Combining with Equation (\ref{eq:canto2}),
the fractional contribution of this inner region to the total drag force is therefore
\begin{equation}
\frac{F_{\rm no-jet}(<R_{\rm max})}{F_{\rm no-jet}(<\infty)}
=\frac{ \ln \left(  2.4 R_{\rm max}/R_{\rm acc}   \right)}
{\ln(3.6 h^{3}\mathcal{M}^{2}q^{-1})}.
\end{equation}
For $R_{\max}=15R_{\rm acc}$, $\mathcal{M}=8$ \citep[as in the simulations of][]{li20}, 
$h=0.03$ and $q=10^{-7}$, corresponding to a $1M_{\odot}$ black hole orbiting
a $10^{8}M_{\odot}$ black hole, this ratio is $0.33$. 
If the jet reduces this inner contribution by a factor of $0.4$, the resulting 
reduction in the total drag force is $\simeq 20\%$. 
Therefore, accretion-powered jets with parameters comparable to those
explored by \citet{li20} could reduce the strength of the drag force, enhancing the
evolution timescale.

\begin{figure}
  \vspace{0pt}
  \includegraphics[angle=0,width=0.99\columnwidth]{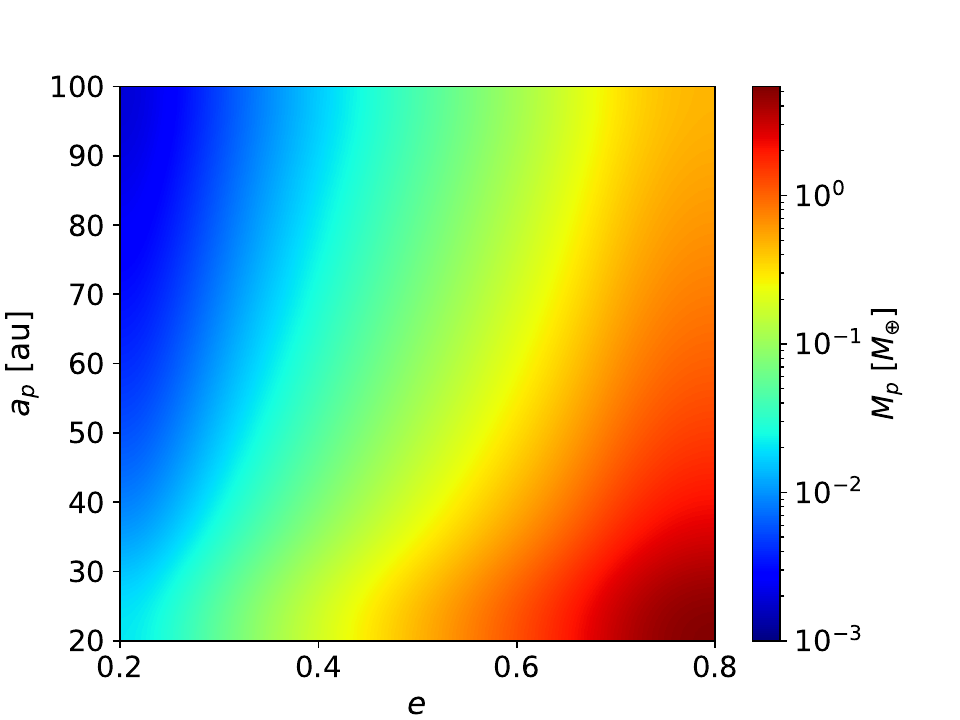}
  \caption{Minimum planetary mass required to satisfy the condition
 $GM_{p}/v_{\rm rel}^{2}>2R_{\rm pl}$ everywhere along 
the orbit, such that treating the planet as a point mass is a reasonable
approximation. We assume a solar-mass central star, a prograde orbit and $h=0.05$.}
  \label{fig:min_mass_point}
\end{figure}

Interestingly, \citet{kaa23} performed
GRMHD simulations in which the jet is launched by a rapidly rotating black hole moving at
$\mathcal{M}=2.45$. They compared the drag for a plasma
$\beta$ of $10$ in two cases: one with a high spin parameter and one
with zero spin and found nearly identical drag forces. This indicates that 
the modification of the drag force is driven primarily by the presence of a magnetic field, rather than by the jet itself. Consistently, in simulations with intermittent jets, the drag force changes only marginally when the jet is active. Given that the strength and large-scale
geometry of magnetic fields in AGN accretion discs are currently essentially unknown, 
the specific role of magnetic fields in determining the drag on supersonically moving
 black holes remains uncertain.

\subsection{Planets in protoplanetary discs}
In addition to gravitational torques, planets embedded in protoplanetary discs may
experience aerodynamic drag when they are in eccentric or inclined orbits \citep{rei12,tey13}, 
as well as heating torques if they release 
luminous energy \citep[e.g.][]{mas17,mas17B,chr23}.
Here we examine the range of parameters for which the orbital evolution is primarily governed by gravitational torques, as estimated above. For clarity, we first consider the effects associated with 
the aerodynamic drag and then turn to the influence of thermal torques.

\subsubsection{The aerodynamic drag}
Our approximation of point masses is valid if the accretion radius $R_{\rm acc}$ is 
larger than several physical (geometrical) radii of the planet $R_{\rm pl}$. 
Numerical simulations of gravitating perturbers with a rigid surface indicate that 
the drag force (including the aerodynamic drag) depends on the nonlinear parameter $\eta$, 
defined as
\begin{equation}
\eta= \frac{1}{2} \frac{\mathcal{M}^{2}}{\mathcal{M}^{2}-1}\frac{R_{\rm acc}}{R_{\rm pl}}
\end{equation}
\citep{thu16,pru24}. 
Consider a planet for which $R_{\rm acc}\simeq 4R_{\rm pl}$ at a
certain point along its orbit. If it moves supersonically, 
then $\eta\simeq 2$. In this case, simulations show that the aerodynamic drag
is negligible compared to the gravitational drag. Moreover, \citet{thu16} report 
a minimum impact parameter of $b_{\rm min}\simeq 1.1R_{\rm acc}$ 
for $\eta\simeq 2$ and a specific heat ratio  
$\gamma=1.2$. Since in a disc with scaleheight $H$, the maximum
effective length is $R_{\rm max}\simeq 3H$ \citep{can13}, Equation (\ref{eq:std}) can be used
to estimate the total drag on an extended object with $\eta=2$,
and to compare it with the prescription adopted in Section \ref{sec:results}.
Substituting $b_{\rm min}$ and $R_{\rm max}$ into
Equation (\ref{eq:std}), we obtain
\begin{equation}
F_{T}\simeq \frac{4\pi \rho_{0}(GM_{p})^{2}}{v_{\rm rel}^{2}} \ln \left(\frac{3H}
{1.1R_{\rm acc}}\right)= \frac{4\pi \rho_{0}(GM_{p})^{2}}{v_{\rm rel}^{2}} \ln\left(\frac{1.6 h^{3}\mathcal{M}^{2}}{q}\right).
\end{equation}
In the last equality, we have used $R_{\rm acc}\simeq 2GM_{p}/(c_{s,\rm adi}^{2}\mathcal{M}^{2})$ 
with $c_{s,{\rm adi}}=\sqrt{\gamma}hv_{K}$ and $\gamma=1.2$. 
For our fiducial values of $h$ and $q$, this force is
$13\%$ lower than that given by Equation (\ref{eq:canto2}) for $\mathcal{M}=3$,
and $11\%$ lower for $\mathcal{M}=8$. Therefore, our results remain robust provided that 
$R_{\rm acc}\gtrsim 4R_{\rm pl}$.

If we demand $R_{\rm acc}\gtrsim 4R_{\rm pl}$ at any point in the orbit, we can
estimate a lower limit to $M_{p}$ by assuming that 
$R_{\rm pl}=(3M_{p}/[4\pi \rho_{\rm pl}])^{1/3}$, where $\rho_{\rm pl}=3$ g cm$^{-3}$
is the mean density of the planet. 
Figure \ref{fig:min_mass_point} shows the lower limit to the planetary
mass, expressed in Earth masses, as a function of the semi-major axis and orbital 
eccentricity, assuming a prograde orbit. Our drag-force formula is expected to be 
quantitatively accurate for non-luminous
planets with masses larger than $\sim 1M_{\oplus}$ over most of the parameter space shown in Figure \ref{fig:min_mass_point}, except for planets with $a_{p}<50$ au and $e>0.6$. In the next 
section we consider the case of luminous planets.

\begin{figure*}
  \vspace{0pt}
\hspace*{-0.0\columnwidth} 
  \includegraphics[angle=0,width=1.0\textwidth]{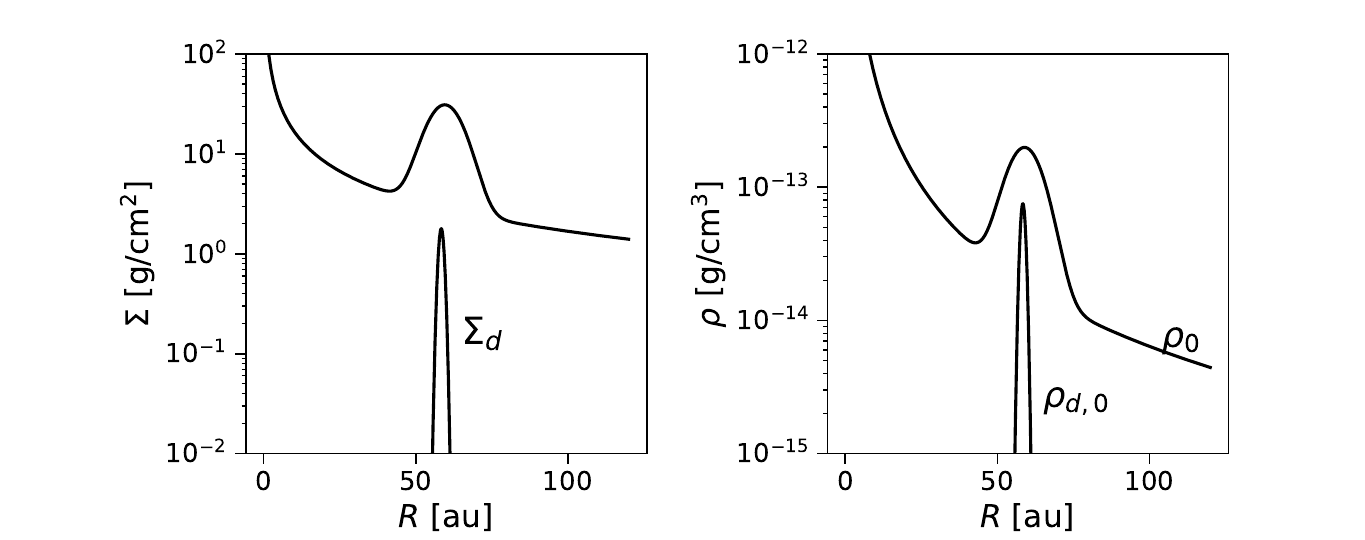}
  \caption{Surface density (left panel) and midplane density (right panel) of the disc model adopted in Section
  \ref{sec:thermal_tq_pl} for the gas (upper curves) and for the dust (lower curves).
}
  \label{fig:surface_density_cgs}
 %This figure is created by mac: Documents/disco6/tex/Rodrigo_Anaya/thermal_torques/Morbidelli_2020/Fig_surface_density_v2.py
\end{figure*}

\begin{figure}
  \vspace{0pt}
  \includegraphics[angle=0,width=0.99\columnwidth]{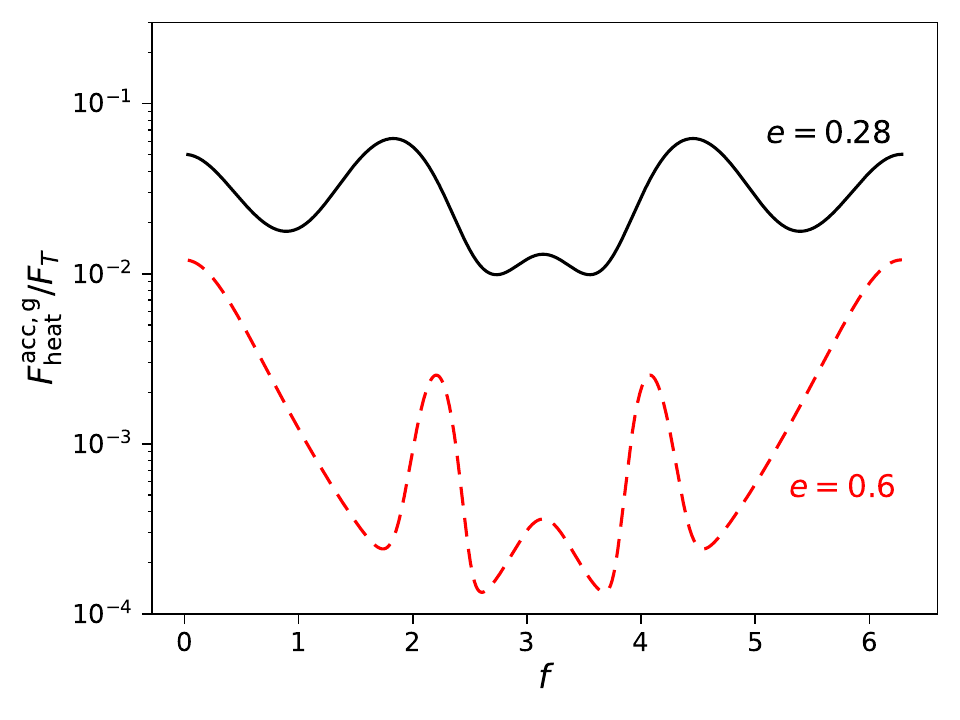}
  \caption{Ratio of the heating force due to gas accretion to $F_{T}$ versus true anomaly $f$
on a prograde planet with $M_{p}=10M_{\oplus}$, $a_{p}=60$ au and two eccentricities. We have adopted $h=0.07$. }
  \label{fig:Fheat_gas_acc}
%This figure is created by mac: disco6/tex/Rodrigo_Anaya/thermal_torques/Morbidelli_2020/Fig_Fheat_gas_acc.py
\end{figure}

\subsubsection{Thermal torques}
\label{sec:thermal_tq_pl}
The structure of the gas in the immediate vicinity of a planet depends on whether 
the planet releases heat 
into its environment and on the thermal diffusivity $\chi$. For non-luminous planets 
on orbits with eccentricities smaller than $h$, thermal diffusion produces overdense,
asymmetric lobes in the planet's coorbital region, enhancing the 
gravitational torque acting on the planet \citep{leg14}. By contrast, when a
planet is sufficiently luminous, again in the low-eccentricity regime, thermal diffusion
leads to the formation of underdense asymmetric lobes. These give rise
to a positive contribution to the torque, the so-called heating torque, which can 
slow down inward migration or even reverse it \citep{2015Natur.520...63B,mas17B}. The luminosity of a planet
is generally attributed to the accretion of pebbles.

In addition, heating torques can also excite the eccentricity of planetary embryos to
values comparable to $h$ \citep[e.g.,][]{chr17,ekl17,cor23,2023MNRAS.520.3286P}, with important dynamical consequences. \citet{vel24} investigated how the thermal torque depends
on eccentricity. For planets with constant luminosity, they found that the thermal torque 
transitions from positive to negative at eccentricities $\sim h$.
\citet{chr23} studied the trapping of low-mass planets in pressure bumps and showed that, for embryos with constant
(supercritical) luminosity and masses below $2M_{\oplus}$,
thermal torques drive the growth of orbital eccentricity, suppressing the positive corotation 
torque that would otherwise trap planets at pressure bumps.
Interestingly, in discs containing a dust ring, the embryo's luminosity increases
when it crosses the ring. Because the luminosity then varies along the epicycle, 
planets can become trapped near the ring, facilitating the accretion of dust from the ring, while maintaining eccentricities comparable to
or smaller than $h$ \citep{pie24,vel24}. %(Pierens \& Raymond 2024; Velasco Romero et al. 2024).
In the following, we evaluate how heat release modifies gas torques for planets on highly eccentric orbits
($e\gtrsim 4h$).

For a planet moving supersonically, one source of heating arises from
the dissipation of gas kinetic energy in the shocked gas downstream of the planet. \citet{mas17} examined
this case and concluded that the resulting heating force is smaller than the standard dynamical friction. 
Here, we recast their argument in quantitative terms. 

For sufficiently low-mass planets ($R_{\rm acc}\ll H$), the luminosity $L_{\rm gas}$
from Bondi-Hoyle-Lyttleton gas accretion is
of order of $GM_{p} \dot{M}_{\rm HL}/R_{\rm acc}$, with $\dot{M}_{\rm HL}=\pi \rho_{0} R_{\rm acc}^{2} 
v_{\rm rel}$ the Bondi-Hoyle accretion rate. The associated heating force is given by
equation (53) in \citet{mas17}
\begin{equation}
\vecF_{\rm heat}^{\rm acc,g} = -\frac{(\gamma-1)GM_{p}L_{\rm gas}}{\chi v_{\rm rel}^{3}}
\ln \Lambda \, \vecv_{\rm rel},
\label{eq:Fheat_gas_acc}
\end{equation}
where $\Lambda=1+\exp\left[-1.96-\ln\left(\frac{r_{\rm min} v_{\rm rel}}{4\chi}\right)\right]$
\citep[see][for a numerical test]{2019MNRAS.483.4383V}. Here $r_{\rm min}$ is the truncation radius adopted in the force calculation, 
which we take to be the larger of the photon mean free path and the accretion radius
$R_{\rm acc}$. Equation (\ref{eq:Fheat_gas_acc}) is valid only under a number of restrictive conditions. 
In particular, the linear regime 
requires $L\ll L_{c}\mathcal{M}$, where $L_{c}$ is defined in
equation (35) of \citet{mas17}. In addition, the planetary mass must
satisfy $M_{p}\leq M_{\rm crit}\equiv c_{s}\chi/G$, ensuring that the thermal diffusion 
timescale is shorter than the acoustic crossing time over the Bondi radius. Furthermore, 
the response time of the heating force must remain shorter than the local dynamical timescale;
this condition implies that the characteristic size of the hot plume be much
smaller than $\mathcal{M}H$ (see Section 6.4 in \citet{mas17}).
Another condition is that the plume size must exceed the photon mean free path, so that the 
diffusion approximation remains valid.

The ratio between the magnitudes of the heating force and the gravitational drag force $F_{T}$  (Equation \ref{eq:canto2}) is 
\begin{equation}
\left|\frac{F_{\rm heat}^{\rm acc,g}}{F_{T}}\right|= \frac{(\gamma-1)\ln\Lambda}{2\ln\left(3.6h^{3}\mathcal{M}^{2}q^{-1}\right)}\left(\frac{GM_{p}}{\chi v_{\rm rel}}\right)
< \frac{(\gamma-1)\ln\Lambda}{2\mathcal{M}\ln\left(3.6h^{3}\mathcal{M}^{2}q^{-1}\right)}.
\label{eq:ratio_gas_acc}
\end{equation}
The last inequality follows from the requirement $M_{p}< c_{s}\chi/G$ for Equation (\ref{eq:Fheat_gas_acc}) to be applicable.
For typical values of $\chi$ in protoplanetary discs, one finds that $\ln\Lambda\lesssim
\ln(3.6h^{3}\mathcal{M}^{2}q^{-1})$. Under these conditions,
the force ratio is bounded by $\lesssim 0.2/\mathcal{M}\simeq 0.4h/e$, where we have set $\gamma=1.4$
and assume a prograde orbit in deriving the last equality.
Consequently, even for prograde planets with masses approaching the critical value $c_{s}\chi/G$
and eccentricities $e\sim 4h$, 
the heating force can partially offset the drag force, reducing its magnitude
by $\sim 20\%$. However, we must ensure that all the conditions outlined in the previous paragraph 
for the applicability of Equation (\ref{eq:Fheat_gas_acc}) are satisfied.

To this end, and to illustrate the relative importance of the heating forces, in the remainder of this section  
we adopt a model expressed in
physical units in order to simplify the presentation. Specifically, for the surface density 
we use a scaled version 
of model 1 at $t=0$, with $M_{\bullet}=1M_{\odot}$, $R_{0}=60$ au, 
$M_{\rm ring}=0.017M_{\odot}$ and $\Sigma_{\rm bg}=2.8$ g cm$^{-2}$ (see the
left panel in Fig. \ref{fig:surface_density_cgs}). For the thermal diffusivity, we adopt
\begin{equation}
\chi = \frac{16(\gamma-1) \sigma T^{3}\mu}{3\rho_{0}^{2} \mathcal{R} \kappa},
\end{equation}
with $\gamma=1.4$, $\sigma$ the Stefan-Boltzmann constant, $\mu$ the mean molar mass ($\mu=2.3$ g/mol), and $\kappa$ is the opacity \citep[e.g.,][]{2017MNRAS.471.4917J}. 
For simplicity, we assume an opacity law of the form
$\kappa = \kappa_{0}T^{2}$, with $\kappa_{0}=5\times 10^{-4}$ cm$^{2}$g$^{-1}$ K$^{-2}$, appropriate for icy grains
\citep{1994ApJ...427..987B}.

We again evaluate 
the ratio $F_{\rm heat}^{\rm acc,g}/F_{T}$ (Equation \ref{eq:ratio_gas_acc})
for a prograde planet of mass $M_{p}=10M_{\oplus}$ along a single orbit with semi-major axis
$a_{p}=R_{0}=60$ au and two eccentricities: $e=0.28$ and $e=0.6$, within the model described above (see 
Figure \ref{fig:Fheat_gas_acc}). The disc temperature is fixed by adopting an
aspect ratio $h=0.07$\footnote{We take $h=0.07$, rather than the nominal 
$h=0.05$, in order to guarantee that the plume size $\simeq \chi/V_{\rm rel}$ is larger than the
photon free mean path $(\kappa \rho_{0})^{-1}$ \citep{mas17}.}. For these parameters, 
$M_{\rm crit}\geq 235M_{\odot}$ at all radii beyond $10$ au. Hence,
the adopted planetary mass remains below $M_{\rm crit}$ over the entire range of orbits considered. 
We find that the heating force 
associated with gas accretion amounts to less than $6\%$
of $F_{T}$ for $e=0.28$, and to less than $1\%$ for $e=0.6$. 
Thus, for $M_{p}\lesssim 10M_{\oplus}$ and $e\geq 4h=0.28$,
the contribution of gas-accretion driven thermal torques is small, although it may
become comparable in magnitude to $F_{T}$ for $M_{p}\gtrsim 50M_{\oplus}$.
However, we find that the condition for a short response time of the heating torque
(i.e. a plume size smaller than $\mathcal{M}H$)
is not satisfied over a large fraction of the orbits.
Consequently, Eq. (\ref{eq:Fheat_gas_acc})
overestimates the heating torque. In fact, we were unable to identify a 
physically plausible model in which
all the required conditions are simultaneously satisfied. Numerical simulations of supersonic bodies embedded in an optically thick medium, in which the heating is self-consistently generated by gas accretion, would be highly valuable for
evaluating the extent to which analytical estimates overpredict the heating torques.

Along with gas, pebble accretion can also contribute to the luminosity of planets. 
To estimate the dust distribution in our disc model, we need the
gas pressure at the midplane
\begin{equation}
P = \frac{\Sigma c_{s}^{2}}{\sqrt{2\pi} H}.
\end{equation}
In our model, $P$ in dyn cm$^{-2}$ can be written as
\begin{equation}
    P = \frac{3.2}{R^{3}} + 1.53\times 10^{-4}\exp\left(-\frac{(R-R_{\rm max})^{2}}{2w^{2}}\right),
\end{equation}
where $R$ is in au, $R_{\rm max}=58.4$ au, and $w=5.8$ au. The steady-state distribution of dust
is the result of the balance between the diffusion and the drag forces of the dust with the gas
\citep{dul18}. 
The radial distribution of dust in the bump is 
\begin{equation}
\Sigma_{d}(R) = \Sigma_{d,{\rm max}}\exp\left(-\frac{(R-R_{\rm max})^{2}}{2w_{d}^{2}}\right),
\end{equation}
where $w_{d}=w/(\sqrt{1+{\rm St}/\alpha})$, with $\rm{St}$ is the Stockes number of dust, $\alpha$
the turbulent parameter, and
\begin{equation}
\Sigma_{0,{\rm max}}= \frac{M_{d}}{(2\pi)^{3/2}R_{\rm max} w_{d}},
\end{equation}
where $M_{d}$ is the dust mass in the ring, which we assume is $M_{d}=10^{-2}M_{\rm ring}$.
The volume density of dust at the midplane is
\begin{equation}
    \rho_{d,0}(R) = \frac{\Sigma_{d}}{\sqrt{2\pi} H_{d}},
\end{equation}
with $H_{d}=H/\sqrt{1+{\rm St}/\alpha}$ \citep{dul18}. Figure \ref{fig:surface_density_cgs} shows the
surface and midplane densities of gas and dust components.

\begin{figure}
  \vspace{0pt}
  \includegraphics[angle=0,width=0.99\columnwidth]{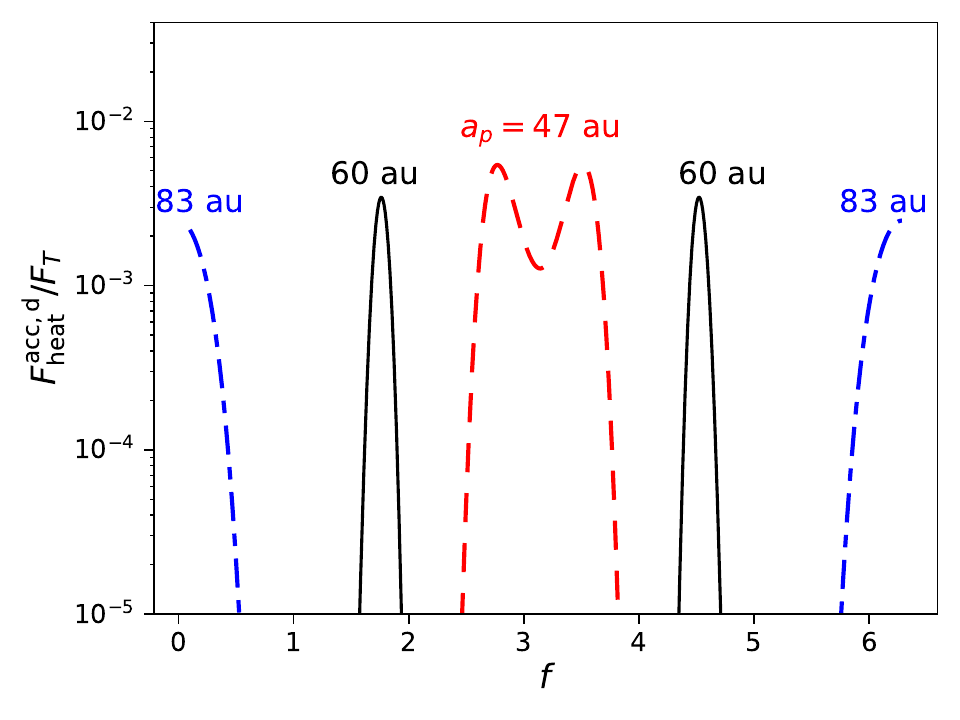}
  \caption{Ratio of the heating force due to pebble accretion to $F_{T}$ as a function of true anomaly $f$
for a prograde planet with $M_{p}=100M_{\oplus}$ and $e=0.28$, shown for three different values of $a_{p}$, as labeled on each curve. }
  \label{fig:Fheat_dust_acc}
%This figure is created by mac: disco6/tex/Rodrigo_Anaya/thermal_torques/Morbidelli_2020/ratio_Fdust_Fg/Fig_Fheat_dust_acc_v2.py
\end{figure}

The luminosity due to the accretion of pebbles is 
\begin{equation}
    L_{d}= \frac{GM_{p} \dot{M}_{d}}{R_{\rm pl}}+ \frac{\dot{M}_{d}v_{\rm rel}^{2}}{2},
\end{equation}
with $\dot{M}_{d}=\pi (1+2\Theta)\rho_{d,0} R_{\rm pl}^{2} v_{\rm rel}$. Here
$\Theta\equiv v_{\rm esc}^{2}/(2v_{\rm rel}^{2})$ is the Safronov number accounting
for gravitational focusing, and $v_{\rm esc}^{2}\equiv 2GM_{p}/R_{\rm pl}$ the escape velocity from the planet. Finally,
the heating force by dust accretion is given by Equation (\ref{eq:Fheat_gas_acc}), with $L_{\rm gas}$
replaced by $L_{d}$, and reads as follows
\begin{equation}
F_{\rm heat}^{\rm acc,d}= -\frac{\pi (\gamma-1)(GM_{p})^{2}\rho_{d,0} R_{\rm pl}}{\chi v_{\rm rel}^{2}} (1+2\Theta)(1+\Theta^{-1})\ln\Lambda\, \vecv_{\rm rel}.
\label{eq:F_acc_d}
\end{equation}
The ratio between the heating force by dust accretion and $F_{T}$ is
\begin{equation}
\left|\frac{F_{\rm heat}^{\rm acc,d}}{F_{T}}\right|=\frac{(\gamma-1) \ln \Lambda}{4\ln(3.6h^{3}\mathcal{M}^{2}q^{-1})}\left(2+\frac{3}{\Theta}+\frac{1}{\Theta^2}\right)
\left(\frac{\rho_{d,0}}{\rho_{0}}\right) \left(\frac{GM_{p}}{\chi v_{\rm rel}}\right).
\end{equation}
 Figure \ref{fig:Fheat_dust_acc} shows this ratio for 
a planet of $100M_{\odot}$ and $e=0.28$ on a prograde orbit, for three different values of 
the semi-major axis. We take again $h=0.07$. The maxima
of the ratio occur at different values of $f$ (depending on the semi-major axis), but its magnitude is similar
in all cases, approximately
$(2-6)\times 10^{-3}$. Since this ratio decreases for lower $M_{p}$ and for higher $e$, the change in the gas torque 
due to heat release driven by dust accretion can be neglected for planets with masses
less than $\sim 100M_{\odot}$ and $e\geq 0.28$. 

Note that, in addition to the gas torque, the planet also gravitationally scatters the dust, forming a dust wake that can contribute to the drag force as the planet crosses the dust ring. 
Moreover, pebble accretion transfers linear momentum from the solids to the planet, 
providing an additional contribution to the total drag. In the dynamical friction approximation, the dust torque is expected to 
act approximately in the same direction as $\vecF_{T}$ and thus counteract the effect of the heating force.

In summary, within the adopted model,
the thermal torque from gas accretion
does not exceed $\sim 5\%$ of $F_{T}$ for $M_{p}\lesssim 10M_{\oplus}$ and $e\geq 4h$,
while the torque associated with
dust accretion is small relative to $F_{T}$ for $M_{p}\lesssim 100M_{\oplus}$ (also for $e\geq 4h$).
We reach the same conclusion for a ring peaking at $R_{0}=10$ au in a disc with
$h=0.05$: within the range of $M_{p}$ for which Eqs. (\ref{eq:Fheat_gas_acc}) and (\ref{eq:F_acc_d})
apply, 
thermal torques are at most $\sim 5\%$ of $F_{T}$ and only during the planet's passage through the ring.

Our dynamical friction analysis is restricted to $e\gtrsim 4h$ and therefore
cannot be applied to the regime $e\lesssim 4h$.
It is nevertheless plausible that the eccentricity continues to damp until it 
approaches $e\simeq h$, after which
migration proceeds radially inward, according to the results of \citet{chr23}. Alternatively,
planets may become trapped on orbits in the vicinity of the ring, thereby inhibiting their migration, 
as suggested by \citet{vel24}.
Simulations of low-mass planets initialized with moderate eccentricities, $e\simeq 0.2$,
would therefore be highly valuable. Such calculations would complement existing studies that focus on nearly circular initial conditions and would help clarify the transition between the high- and low-eccentricity regimes.

\section{Summary and conclusions}
\label{sec:summary}
Dust substructures like arcs, rings and spiral arms have been observed in protoplanetary discs through
high-resolution interferometric imaging. Some models suggest that these
substructures occur by dust trapping in gas pressure bumps. Planetary migration in discs
with pressure bumps or cavities has often been investigated under the assumption
of circular orbits. Yet, several dynamical processes can
excite planetary eccentricities, including planet-planet scattering events
\citep[see][for a review]{paa23}. Streamers feeding
the disc may not only generate spiral arms and ring-like features but 
can also impart torques that drive a global eccentric mode. This mode can secularly
couple to an embedded planet and sustain a finite orbital eccentricity.

In a different astrophysical context, compact objects in the nuclear stellar cluster 
surrounding AGNs can become embedded in the accretion disc once their orbits intersect it. Such captures can occur over the full range of orbital eccentricities.

Motivated by these processes, we investigated the orbital 
evolution of low-mass perturbers embedded in a gaseous 
disc containing a prominent axisymmetric, ring-like overdensity, using the 
local approximation (or dynamical friction approach). We examined
both prograde and retrograde orbits relative to the disc's rotation.
In the prograde case we restricted our analysis to eccentricities above  
$4h$, ensuring that the perturber's motion relative to the disc
remains supersonic. In the retrograde case, the motion is supersonic for any value
of the eccentricity.

We computed maps of the characteristic timescales $t_{a}$ and $t_{e}$ for both prograde and retrograde orbits. As expected, if $t_{\nu}$, the timescale over which the ring undergoes
viscous radial spreading, is $\ll {\rm min}(t_{a},t_{e})$, the ring vanishes
before it can appreciably affect the perturber's orbit. 

In cases where $t_{\nu}\gtrsim {\rm min}(t_{a},t_{e})$, our analysis reveals the following key features
of the orbital evolution of perturbers.
For prograde orbits, the rates of change $|t_{a}|^{-1}$ and $|t_{e}|^{-1}$
reach their maximum values when the apocentre or pericentre lies at or near the ring's
density peak. In all prograde orbits, the eccentricity is invariably damped. Yet,
quite remarkably, we identified a region where
the presence of the ring reverses the usual trend and drives migration outward. 
Prograde orbits 
intersecting the circle $R=R_{0}$ (the location of the ring's density maximum), 
progressively circularize while their semi-major axes converge 
toward $R_{0}$. Consequently, perturbers accumulate there,  giving rise to a ring 
superimposed on the gaseous ring. Moreover, orbits whose
apocentres or pericentres lie at or near $R=R_{0}$, evolve while keeping these points close
to $R_{0}$. In turn, perturbers confined to such 
orbits experience the highest mass accretion rates.

Our analysis is restricted to eccentricities $e\gtrsim 4h$. At lower eccentricities, 
the convergence of the planetary orbits toward the ring may be disrupted by thermal 
torques that can maintain eccentricities at levels of $\sim h$ and thereby may suppress corotation torques,
as shown by \citet{chr23}. 
Alternatively, planets may become trapped on orbits that progressively deplete the dust rings, as 
proposed by \citet{pie24} and \citet{vel24}. Conducting simulations of low-mass planets with initial 
eccentricities around $e\simeq 0.2$ could provide useful insights.

For retrograde orbits, $t_{a}$ and $t_{e}$ are larger than for prograde orbits, making
the condition $t_{\nu}\gtrsim {\rm min}(t_{a},t_{e})$ correspondingly harder to satisfy.
The retrograde case exhibits features that are distinct from those of the prograde case.
Migration invariably proceeds inward and, thus, there is no region where perturbers
can accumulate. However, the eccentricity may be either excited
or damped, depending on $(e, a_{p})$, with the necessary condition for
damping being $a_{p}>R_{0}$. Generally, orbits with $R_{0}<a_{p}<1.8R_{0}$ and
$e<0.8$ experience an
initial phase in which eccentricity is damped or only mildly excited.
As $a_{p}$ decreases below $R_{0}$, the eccentricity begins to grow. For the ring profiles
considered in this work, the eccentricity growth rate in this phase is comparable to, or 
even exceeds, the migration rate. Yet, the eccentricity is not large enough for the orbits
to reach the ring any longer, so the orbits become confined within it.

Our analysis relies on the assumption that the torque and power are accurately
captured by the local, or dynamical friction, approximation. This
approximation is generally valid as long as the Bondi-Hoyle-Lyttleton accretion
radius is smaller than the disc's scale height \citep{san18,san19}. Nonetheless, it would be
highly valuable to test our predictions with numerical simulations. Ideally, such
simulations should be three-dimensional in order to resolve the wake across
the disc's vertical thickness, and are therefore computationally demanding.

We modelled the perturber as a point-mass accretor and assumed an
isothermal disc. However, we also examined 
additional physical processes that may be relevant in specific astrophysical scenarios. For compact objects 
on (supersonic) eccentric orbits embedded in AGN accretion discs, previous radiation-hydrodynamic
simulations indicate that 
radiative feedback does not significantly modify the drag force.
The impact of jets on reducing the drag force remains highly uncertain for supersonic
objects, since jet launching is likely 
magnetically driven and therefore depends on magnetic field properties that 
are poorly constrained observationally in AGN discs.

For planetary cores and planets embedded in protoplanetary discs, we relaxed the point-mass 
approximation and evaluated the aerodynamic drag, together with the contribution from thermal torques.
We find that the aerodynamic drag remains modest 
($\lesssim 15\%$) compared to the dynamical friction force
for masses $\geq 1M_{\oplus}$. Across the parameter space where the adopted 
analytical expressions apply, thermal torques contribute less than $5\%$ of $F_{T}$ for the models examined 
in this study.

\section*{Acknowledgments}
The paper benefited greatly from the constructive suggestions and encouragement provided by the referee.

\section*{Data Availability}
This study does not involve the use or production of original data. 
The Python scripts used to perform the calculations will be shared on reasonable request 
to the corresponding author.

\bibliographystyle{mnras}
\bibliography{viscous}
% Don't change these lines
\bsp	% typesetting comment
\label{lastpage}
\end{document}